\documentclass[%
 aip,
 amsmath,amssymb,
 reprint,%
]{revtex4-1}
\usepackage{graphicx}
\usepackage{subfigure}
\usepackage{dcolumn}
\usepackage{bm}
\usepackage{color}
\usepackage[T1]{fontenc}
\usepackage{mathptmx}
\usepackage{multirow}
\usepackage{array}
\usepackage{float}
\usepackage{url}
\usepackage{titlesec}
\titleclass{\subsubsubsection}{straight}[\subsection]
\newcounter{subsubsubsection}[subsubsection]
\renewcommand\thesubsubsubsection{\thesubsubsection.\alph{subsubsubsection}}
\titleformat{\subsubsubsection}
 {\normalfont\normalsize\bfseries}{\thesubsubsubsection}{1em}{}
\titlespacing*{\subsubsubsection}
{0pt}{3.25ex plus 1ex minus .2ex}{1.5ex plus .2ex}

\begin{document}
\preprint{AIP/123-QED}

\title{Viscous effects on morphological and thermodynamic non-equilibrium characterizations of shock-bubble interaction}

\author{Dejia Zhang}
 \affiliation{State Key Laboratory for GeoMechanics and Deep Underground Engineering, China University of Mining and Technology, Beijing 100083, P.R.China}
 \affiliation{National Key Laboratory of Computational Physics, Institute of Applied Physics and Computational Mathematics, P. O. Box 8009-26, Beijing 100088, P.R.China}
 \affiliation{National Key Laboratory of Shock Wave and Detonation Physics, Mianyang 621999, China}

\author{Aiguo Xu}
 \thanks{Corresponding author: Xu\_Aiguo@iapcm.ac.cn}
\affiliation{National Key Laboratory of Computational Physics, Institute of Applied Physics and Computational Mathematics, P. O. Box 8009-26, Beijing 100088, P.R.China}
\affiliation{HEDPS, Center for Applied Physics and Technology, and College of Engineering, Peking University, Beijing 100871, China}
\affiliation{State Key Laboratory of Explosion Science and Technology, Beijing Institute of Technology, Beijing 100081, China}

\author{Yanbiao Gan}
\affiliation{Hebei Key Laboratory of Trans-Media Aerial Underwater Vehicle, School of Liberal Arts and Sciences, North China Institute of Aerospace Engineering, Langfang 065000, China}

\author{Yudong Zhang}
\affiliation{School of Mechanics and Safety Engineering, Zhengzhou University, Zhengzhou 450001, P.R.China}

\author{Jiahui Song}
\affiliation{School of Aerospace Engineering, Beijing Institute of Technology, Beijing, 100081, P.R.China}
 \affiliation{National Key Laboratory of Computational Physics, Institute of Applied Physics and Computational Mathematics, P. O. Box 8009-26, Beijing 100088, P.R.China}
 \affiliation{National Key Laboratory of Shock Wave and Detonation Physics, Mianyang 621000, China}

\author{Yingjun Li}
\thanks{Corresponding author: lyj@aphy.iphy.ac.cn}
\affiliation{State Key Laboratory for GeoMechanics and Deep Underground Engineering, China University of Mining and Technology, Beijing 100083, P.R.China}%

\date{\today}

\begin{abstract}

A two-fluid discrete Boltzmann model (DBM) with a flexible Prandtl number is formulated to study the shock-bubble interaction (SBI).
This paper mainly focuses on the viscous effects on morphological and Thermodynamic Non-equilibrium (TNE) characterizations during the SBI process.
Due to the rapid and brief nature of the SBI process, viscosity has a relatively limited influence on macroscopic parameters but significantly affects the TNE features of the fluid system.
Morphologically, viscosity affects the configuration of the vortex pair, increases both the amplitudes of gradients of average density and average temperature of the fluid field, and reduces circulation of the bubble.
As a higher viscosity fluid absorbs more energy from the shock wave, it leads to an increase in both the proportion of the high-density region and the corresponding boundary length for a fixed density threshold.
The spatiotemporal features of TNE quantities are analyzed from multiple perspectives.
The spatial configuration of these TNE quantities exhibit interesting symmetry, which aids in understanding the way and extent to which  fluid unit deviating from equilibrium state.
Theoretically, viscosity influences these TNE quantities by affecting the transport coefficients and gradients of macroscopic quantity.
Meanwhile, the viscosity increases the entropy production rate originating from the non-organized momentum flux mainly through amplifying the transport coefficient, and enhances the entropy production rate contributed by the non-organized energy flux by raising the temperature gradient.
These multi-perspective results collectively provide a relatively comprehensive depiction of the SBI.

\end{abstract}

\maketitle

\section{\label{sec:level1} Introduction}

The physical scenarios of shock-bubble interaction (SBI) are ubiquitous in both natural phenomena and engineering applications. \cite{Ranjan2011ARFM,Ranjan2007PRL,Liu2022JFM,Yu2021PRF,
Zou2016POF,Fan2022CNF,Liang2018SCPMA,Rawat2023POF}
For example, in astrophysics, the Puppis A supernova remnant interacts with a complex system of interstellar clouds. \cite{Hwang2005}
In combustion systems, the shock wave ignites the mixture bubble composed of $\rm{H_2}$ and $\rm{O_2}$. \cite{Diegelmann2017CNF}
In inertial confinement fusion, laser-induced shock wave impacts the isolated defects bubble inside the capsule, aggravating hydrodynamic instability. \cite{Liu2023POP}
In the field of medicine, shock waves are directed into the body and fragment the kidney stones inside patient. \cite{Lingemen2009NRU,Leighton2013PRSA}
A fundamental setup for studying SBI involves a spherical bubble accelerated by a planar shock wave.
 The deformation of the bubble is influenced by numerous physical factors, including the type and strength (expressed as Mach number, Ma) of the incident shock, the initial bubble shape, the boundary types of fluid field, the density ratio between the bubble and the ambient gas (Atwood number), the specific-heat ratio, the viscosity, and the heat conduction, etc.
Due to the significance of SBI in engineering applications, substantial efforts have been dedicated to unraveling its evolutionary mechanisms.
Researchers have employed variety of approaches, primarily encompassing theoretical methods, \cite{Samtaney1994JFM,Yang1994JFM} experimental investigations, \cite{Ranjan2007PRL,Layes2003PRL,Haas1987JFM,
Zhai2019POF,Kitamura2022POF} and numerical simulations \cite{Picone1988JFM,Zou2015SCPMA,Sha2015ActaPS,
Zhu2017POF,li2019POF,Chen2021POF,Igra2023POF,Guan2022POF}.
Among these, Samtaney \emph{et al.}\cite{Samtaney1994JFM} provided the analytical expressions for circulation $\Gamma$ which are within and beyond the regular refraction regime.
Ding \emph{et al.}\cite{Ding2017JFM,Ding2018POF} conducted experimental and numerical investigations into the effects of initial interface curvature on the interaction between planar shock waves and heavy/light bubbles.
Additionally, other scenarios, such as bubbles impacted by the converging shock wave, \cite{Si2014LPB} spherical/cylindrical bubbles interact with planar shock wave under re-shock conditions, \cite{Si2012POF,Zhai2014JV} have also been explored.

Numerical simulations of SBI can be categorized into three types based on their theoretical foundation: macroscopic, mesoscopic, and microscopic methods.
In previous studies, the traditional macroscopic modeling method, founded on the continuum hypothesis (or equilibrium and near-equilibrium hypothesis), has been widely employed.
Such macroscopic fluid models are often represented by the Euler equations and Navier-Stokes (NS) equations, wherein the former assumes equilibrium, and the latter assumes near-equilibrium.
The hydrodynamic equations in the traditional macroscopic modeling method only describe the hydrodynamic behaviors corresponding to the conservation laws of mass, momentum, and energy.
However, with increasing the degrees of non-equilibrium and non-continuity, more appropriate hydrodynamic equations refer to the Extended Hydrodynamic Equations (EHEs) which encompass not only the evolution equations of conserved kinetic moments but also the most relevant non-conserved kinetic moments of the distribution function. \cite{Zhang2022POF}
The traditional macroscopic model describes the SBI process from a macroscopic perspective, primarily encompassing the fields of density, temperature, velocity, and pressure, which has been instrumental in advancing our understanding of the physical aspects of SBI.
For example, Ding \emph{et al.}\cite{Ding2017JFM,Ding2018POF} demonstrated a good agreement of interface structure between the numerical results obtained from the compressible Euler equations and experimental data.
Zou \emph{et al.}\cite{Zou2015SCPMA} investigated the Atwood number effects and the jet phenomenon caused by the shock focusing through the multi-fluid Eulerian equations.
In contrast, a smaller portion of SBI research utilizes the mesoscopic method, such as the Direct Simulation Monte Carlo method. \cite{Zhang2019CNF}

Two challenges are encountered in the previous SBI numerical studies.
(i) Most of these studies describe mainly the flow morphology and SBI process from a macroscopic view.
They are concerned more with dynamic processes such as bubble deformation, interface motion, vortex motion, mixing degree, etc.
These physical quantities are helpful for understanding the flow morphology during the SBI process but are far from being sufficient.
However, with increasing the non-continuity and Thermodynamic Non-Equilibrium (TNE), the complexity of system behaviors increases sharply.
To ensure that the ability to describe the system does not decrease, it is necessary to incorporate additional physical quantities, such as TNE parameters, to fully characterize its state and behavior.
Many studies have emphasized the importance of investigating TNE behaviors to gain insights into the kinetic processes. \cite{Gan2011PRE,Lai2016PRE,Lin2017PRE,Lin2021PRE,
Chen2018POF,Gan2019FOP,Chen2022PRE,Li2022CTP,
Shan2022JMES,Liu2022JMES,Zhang2021POF,Chen2022FOP,
Zhang2023CAF,Chen2021FOP,Su2022CTP,ZhangYD2023POF}
Among these, Zhang \emph{et al.}\cite{Zhang2023CAF} studied the specific-heat ratio effects on kinetic processes of SBI from multiple perspectives.
(ii) The viscous effects on small-scale structures and kinetic features of SBI need further investigation.
Research conducted by Zhang \emph{et al.} \cite{Zhang2021POF} presented the significant role of viscosity in material mixing in single-mode Rayleigh–Taylor (RT) system.
Zhang \emph{et al.} \cite{Zhang2019CNF} demonstrated that the viscous effects lead to the disappearance of some typical phenomena on the reacting shock-bubble interaction.
The bulk viscosity associated with the viscous excess normal stress, including different physical properties of diatomic and polyatomic gases, significantly changes the flow morphology and results in complex wave patterns, vorticity generation, vortex formation, and bubble deformation. \cite{Singh2021POF}
Moreover, studying the viscous effects on kinetic features, particularly on TNE features,  is crucial for understanding the fundamental mechanism of viscous effects.

\begin{figure*}[htbp]
\center\includegraphics*
[width=0.8\textwidth]{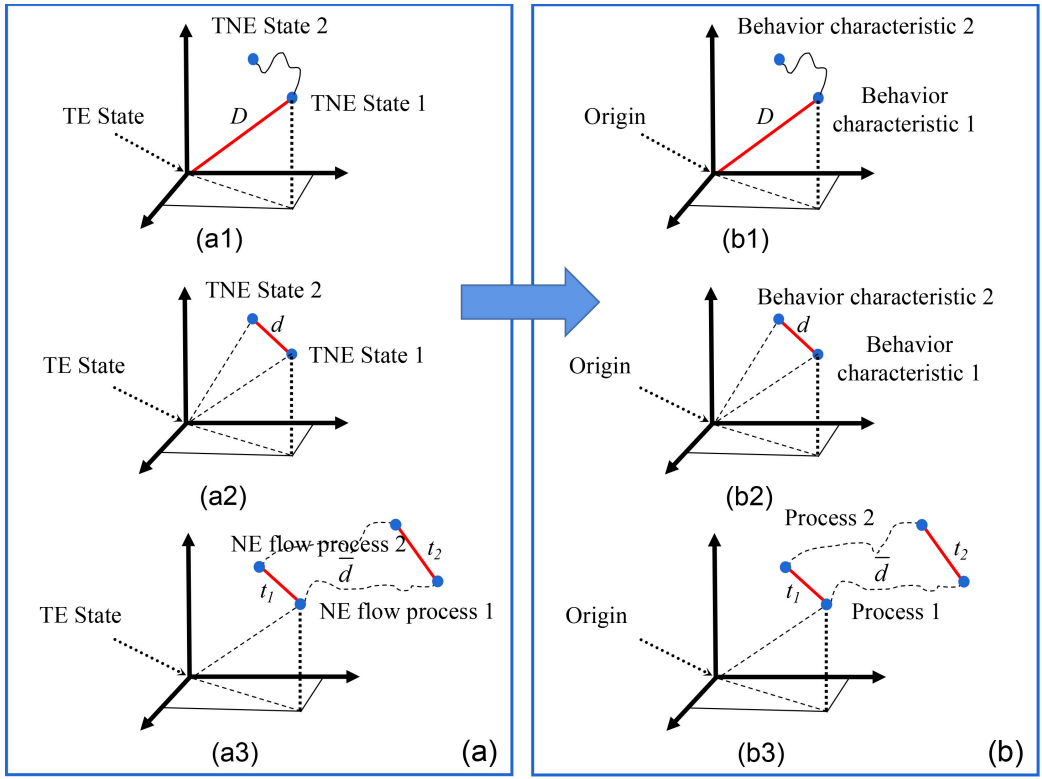}
\caption{ Schematic space opened by independent components of TNE quantities, where ``TE'' represents thermodynamic equilibrium and ``NE'' indicates non-equilibrium. }
\label{fig00}
\end{figure*}

To further investigate the aforementioned inadequacies, we can employ the recently proposed discrete Boltzmann method/model/modeling
\footnote[1]{The DBM can be interpreted as discrete Boltzmann method/model/modeling according the specific context.} (DBM). \cite{Zhang2023CAF,Zhang2022POF,Xu2022CMK,Xu2023arXiv,
Gan2022JFM,Zhang2022AIP,Zhang2017FOP,Zhang2019Matter,
2021XuACTAA,2021XuCJCP,2021XuACTA}
Historically, there are two complementary branches for Lattice Boltzmann Method (LBM) research. \cite{Xu2022CMK,Zhang2023CAF,2001-Succi-Boltzmann,Higuera1989EL,Benzi1992PR,Orazio2003ICCS}
One is to construct physical models that connect microscopic and macroscopic descriptions.
The other aims to develop a new kind of scheme that numerically solves the various partial differential equation(s).
The current DBM is evolved from the physical modeling branch of LBM with some additions and some abandons.
\emph{The DBM serves as a physical modeling and the complex physical field analysis method.}
Its tasks mainly include two aspects: (i) Capturing the main features of the problems to be studied.
In addition to mass, momentum, and energy conservation moments, the DBM considers more relevant non-conserved kinetic moments that describe the main TNE behaviors of the system.
With the increase of TNE, more non-conserved kinetic moments should be required to ensure the non-significant decrease of the system state and behavior description function.
Through the Chapman-Enskog (CE) multiscale analysis, the kinetic moments describe the system state and features can be quickly determined.
It should be noted that in complex systems, each kinetic moment represents one perspective and complex systems require multi-perspective research.
The result from these perspectives together constitutes a relatively complete description of the system.
(ii) Trying to extract more valuable physical information for massive data and complex physical fields.
Based on the non-equilibrium statistical physics, the DBM uses the non-conservative moments of ($f-f^{eq}$), i.e., the TNE quantities, to describe how and how much the system deviates from the thermodynamic equilibrium state and to check corresponding effects due to deviating from the thermodynamic equilibrium. \cite{Xu2022CMK}
The TNE quantities open a high-dimensional phase space, and this phase space, along with its sub-space, provide an intuitive geometric framework for understanding complex behaviors.
Figure \ref{fig00}(a) shows the schematic space opened by independent components of TNE quantities.
In the phase space, the origin represents the thermodynamic equilibrium state, and a state point indicates a specific TNE state.
The distance $D$ is used to describe the degree of a specific TNE state deviating from the equilibrium state.
The distance $d$ between two state points can define as the difference between two TNE states, and its reciprocal, $1/d$, can define as the similarity between the two TNE states.
The mean distance ($\overline{d}$) during the kinetic process can roughly characterize the difference between the two corresponding kinetic processes.
The reciprocal of mean distance, $1 / \overline{d}$, can describe the process similarity.
As shown in Fig. \ref{fig00}(b), the phase space description methodology can be extended to any set of system characteristics.
Many kinds of TNE quantities can be defined according to the research requirement, and each TNE characteristic quantity describes the TNE behaviors from its own perspective.

Other analysis methods, such as the morphological analysis method based on the Minkowski measures, \cite{Xu2009PD} and the description of tracer particle method, \cite{Zhang2021POF} are also coupled in DBM.
These analysis approaches together constitute a relatively complete description for complex physical field.
Summarily, DBM modeling surpasses the traditional macroscopic modeling in at least two aspects: (i) Extended the description ability for non-equilibrium flows.
The physical function of DBM corresponds to the EHEs which considers not only the evolution of conserved kinetic moments but also the evolution of the most relevant non-conserved moments.
That allows DBM to more accurately describe the higher TNE flows.
(ii) Provided a set of analysis methods for complex physical field.
In numerical simulation research, the NS model is only responsible for physical modeling before simulation and do not address the analysis of the complex physical fields after simulation.
While the DBM is responsible for both pre-simulation and post-simulation.

In the following part, section \ref{Model construction} outlines the discrete Boltzmann method construction process which includes the coarse-grained modeling part and the analysis scheme construction part, corresponding to steps (i) and (iii) in Fig. \ref{fig0}(a).
Section \ref{Numerical simulations} presents the numerical results and Sec. \ref{Conclusions} makes the conclusion.

\section{Discrete Boltzmann method construction process}\label{Model construction}

\begin{figure*}[htbp]
\center\includegraphics*
[width=0.9\textwidth]{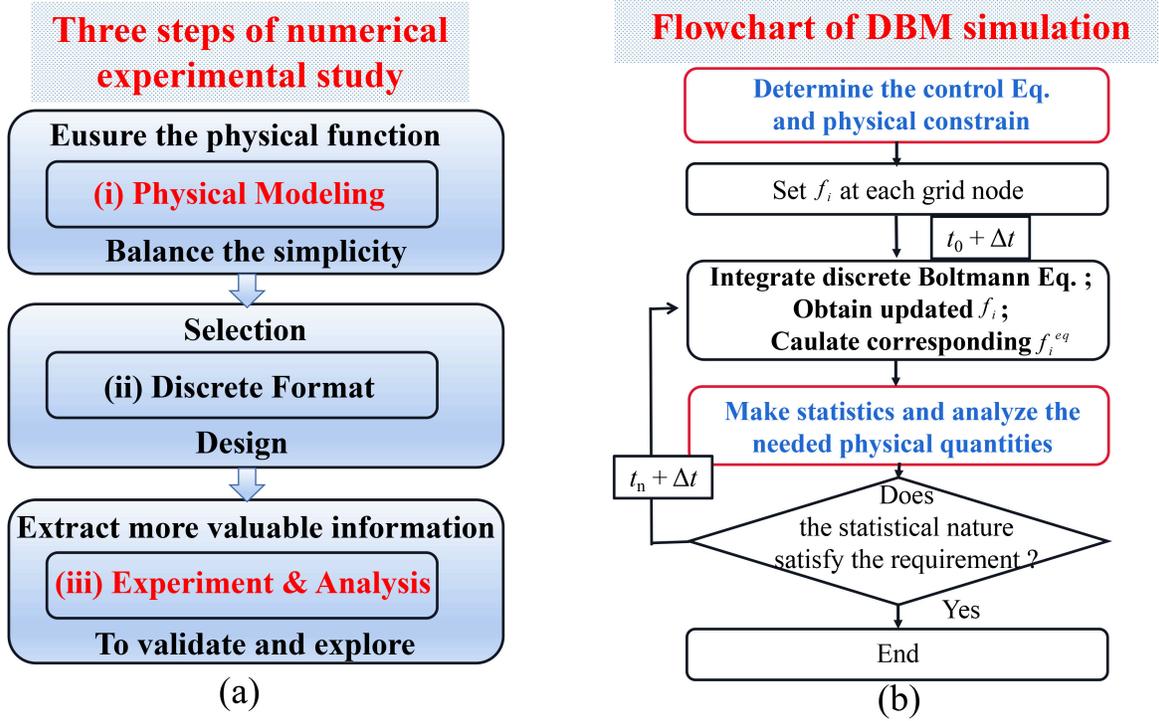}
\caption{ (a) Three steps of numerical study. The DBM works for part (i) and (iii). (b) Flowchart of DBM simulation. }
\label{fig0}
\end{figure*}

As shown in Fig. \ref{fig0}(a), the numerical experimental research mainly includes three parts: (i) physical modeling, (ii) selection/design of discrete format, and (iii) numerical experimental study and analysis.
The DBM is a kind of physical modeling complex physical field analysis method, so it only works for the parts (i) and (iii).
The discrete format researches in step (ii)  are not included in DBM and DBM is just the users of discrete format.
Therefore, the complete interpretation of DBM is Discrete Boltzmann \emph{modeling and analysis} Method.
Summarily, there are mainly two tasks for DBM: (i) for the specific physical problems to be investigated, the DBM should ensure the rationality of the theoretical model and give consideration to the simplicity, (ii) for the massive data and complex physical fields, the DBM is aims to try extract more valuable and helpful information.
Figure \ref{fig0}(b) is the flowchart of DBM simulation.
It shows the two steps (the red boxes in the figure) that are focused in DBM numerical simulation: (i) for a specific physical problem, determining the control equation and physical constrain, and (ii) making statistics and analyzing the needed physical quantities.
Apparently, the two steps correspond to physical modeling and complex physical field analysis in numerical experimental research, respectively.

Physically, to describe the physical image of SBI, and to investigate the viscous effects on SBI, we can choose a first-order two-fluid DBM with flexible Prandtl (Pr) numbers, where ``first-order'' means only the first order of TNEs are considered in the modeling process.\cite{Zhang2023CAF}
Therefore, in the following part, we take this first-order two-fluid DBM as example, and demonstrate in detail its physical modeling steps.
The method construction process of DBM contains two parts: the coarse-grained physical modeling before simulation and the complex physical field analysis methods after simulation.
Among them, the coarse-grained physical modeling includes simplification of collision operator and discretizing of the particle velocity space.

\subsection{ Coarse-grained modeling part }

\subsubsection{Simplification of collision operator}

To facilitate the solution of the Boltzmann equation, the original complex collision operator should be simplified.
A common method, through introducing a local equilibrium distribution function $f^{eq}$ and writing the collision operator into the linearized form, is presented by Bhatnagar, Gross and Krook (BGK). \cite{Bhatnagar1954BGK}
In the simplification step, it requires that the kinetic moments described the physical problems cannot change their values after simplification, i.e., $\int Q(f, f^{'})\Psi(\mathbf{v}) d\mathbf{v}$ = $\int -\frac{1}{\tau}(f-f^{eq})\Psi(\mathbf{v}) d\mathbf{v}$, where $\Psi = [1,\mathbf{v},\mathbf{v}\mathbf{v},\mathbf{v}\mathbf{v}\mathbf{v},\cdots]^{T}$ represents the concerned kinetic moments.
It should be noticed that the original BGK model only describes the situations correspond to the quasi-static and quasi-equilibrium.
\emph{The currently used BGK-like models for non-equilibrium flows are the modified versions that incorporate the mean-field theory description}. \cite{Xu-KouShare,Gan2022JFM,Xu2022CMK}
To realize a adjustable $\Pr$ number, the Ellipsoidal Statistical Bhatnagar-Gross-Krook (ES-BGK) Boltzmann equation is adopted in this work. \cite{Zhang2017FOP,1966ES,Zhang2020POF}
Up to this step, the Boltzmann equation in ES-BGK form is obtained, i.e.,
\begin{equation}
\frac{\partial f}{\partial t}+\bm{v}\cdot\frac{\partial f}{\partial \bm{r}}=-\frac{1}{\tau}(f-f^{ES})
\label{Eq.Boltzmann-equation}
,
\end{equation}
where $\bm{v}$, $\bm{r}$, $t$, and $\tau$ represent the particle velocity, particle position, time, and relaxation time, respectively.
The ES distribution function $f^{ES}$ is
\begin{equation}
\begin{aligned}
f^{ES}=\frac{\rho}{2\pi \sqrt{\left|\lambda_{\alpha\beta}\right|}}\times \exp[-\frac{1}{2}\lambda_{\alpha \beta}^{-1}(v_{\alpha}-u_{\alpha})(v_{\beta}-u_{\beta})]
,
\label{Eq.fES}
\end{aligned}
\end{equation}
where $\rho$ is the mass density and $u_{\alpha}$ ($u_{\beta}$) represents the flow velocity in $\alpha$ ($\beta$) ($\alpha,\beta$ = $x$ or  $y$) direction.
The modified term is $\lambda_{\alpha\beta}=RT\delta_{\alpha\beta}+\frac{b}{\rho}\Delta^{*}_{2,\alpha\beta}$, with $b$ a flexible parameter adjusting the $\Pr$ number, i.e., $\Pr=1/(1-b)$.
When $b=0$, $f^{ES}$ is reduced to $f^{eq}$ and the ES-BGK form is simplified to the BGK form.

\subsubsection{Discretizing of particle velocity space}

For simulation, the continuous Boltzmann equation should be discretized in its velocity space.
In this step, the continuous kinetic moments should be transferred into summation form.
The DBM must ensure that the reserved kinetic moments describe the system behaviors need to keep their values after discretizing the velocity space, i.e., $\int f \Psi '(\mathbf{v}) d\mathbf{v}=\sum_{i} f_i \Psi '(\mathbf{v}_i)$.
Because the distribution function $f$ can be expressed by $f^{eq}$, so the reserved kinetic moments of $f^{eq}$ should keep their values, i.e., $\int f^{eq} \Psi ''(\mathbf{v}) d\mathbf{v}=\sum_{i} f_i^{eq} \Psi ''(\mathbf{v}_i)$, where $\Psi '$ and $\Psi ''$ represent the conserved kinetic moments.
The conserved kinetic moments depend on specific physical problems and they give the most necessary physical constrains for the discrete way in discretizing the velocity space.
The discrete Boltzmann equation in ES-BGK form is
\begin{equation}
\frac{\partial f_{i}}{\partial t}+v_{i\alpha}\cdot\frac{\partial f_{i}}{\partial r_{\alpha}}=-\frac{1}{\tau}(f_{i}-f^{ES}_{i})
,
\end{equation}

In DBM, the CE analysis is used to determine the type and number of the conserved kinetic moments.
Specifically, five kinetic moments are identified when constructing a first-order DBM, i.e., $\mathbf{M}_0^{ES}$, $\mathbf{M}_1^{ES}$, $\mathbf{M}_2^{ES}$, $\mathbf{M}_3^{ES}$, $\mathbf{M}_{4,2}^{ES}$.
Their expressions are provided in the Appendix \ref{sec:AppendixesB}.
They can be obtained by integrating the Eq. (\ref{Eq.fES}) in the particle velocity space $\bf{v}$.
Because the $f^{ES}$ is related to $\Delta_{2,\alpha\beta}^{*}$, it also requires the kinetic moments of $f^{eq}$ in the modeling process, i.e., $\mathbf{M}_0^{eq}$, $\mathbf{M}_1^{eq}$, $\mathbf{M}_2^{eq}$, $\mathbf{M}_3^{eq}$, $\mathbf{M}_{4,2}^{eq}$.
Their expressions can be obtained by setting $b=0$ into $\mathbf{M}_{m}^{ES}$ ($\mathbf{M}_{m,n}^{ES}$).
For convenience, we can write the kinetic moments into a matrix equation
\footnote[2]{Here we consider the kinetic moments of component $\sigma$, where $\sigma$ represents the type of component.
The introduction of two-fluid system can be seen in the following discussions.}, i.e.,
\begin{equation}
\mathbf{C}\cdot\mathbf{f}^{\sigma,ES}=\mathbf{\hat{f}}^{\sigma,ES}
,
\label{Eq.fES}
\end{equation}
and
\begin{equation}
\mathbf{C}\cdot\mathbf{f}^{\sigma,eq}=\mathbf{\hat{f}}^{\sigma,eq}
,
\label{fieq}
\end{equation}
Equations (\ref{Eq.fES}) and (\ref{fieq}) represent the physical constrains imposed by the discretization step in DBM modeling.
According to the number of the reserved kinetic moment, the dimension of $\mathbf{\hat{f}}^{\sigma,eq}$ ($\mathbf{\hat{f}}^{\sigma,ES}$) is $\mathbf{\hat{f}}^{\sigma,eq}=(\hat{f}_{1}^{\sigma,eq},\hat{f}_{2}^{\sigma,eq},\cdots, \hat{f}_{N_{\rm{moment}}}^{\sigma,ES})^T$ [$\mathbf{\hat{f}}^{\sigma,ES}=(\hat{f}_{1}^{\sigma,ES},\hat{f}_{2}^{\sigma,ES},\cdots, \hat{f}_{N_{\rm{moment}}}^{\sigma,ES})^T$], with $N_{\rm{moment}}=13$.
$\mathbf{C}$ represents the matrix of discrete velocity model (DVM).
It should be noted that the DBM retains the use of discrete velocities but does not adhere to the specific discrete format, only gives the most necessary physical constraints for the discrete velocities selection to follow.
In summary, during the coarse-grained physical modeling processes, the DBM offers the necessary physical constrains that should be followed.

According to the CE multiscale analysis, the Boltzmann equation can be reduced to the macroscopic hydrodynamic equations, as detailed in Appendix \ref{sec:AppendixesC}).
It should be noted that recovering the corresponding hydrodynamic equations is only one aspect of the DBM's physical function.
Physically, the physical function of DBM corresponds to the EHEs which not only reserve the conserved kinetic moments evolution equations, but also some of the most closely related non-conserved moments evolution equations.
The modeling method that derives EHEs from the Boltzmann equation is referred to as Kinetic Macroscopic Modeling (KMM) method.
However, the DBM is a kind of Kinetic Direct Modeling (KDM) method.
In DBM, deriving the hydrodynamic equations is used to verify the correctness of physical modeling, the DBM does not need to solve the hydrodynamic equations.

\subsubsection{ Introducing of two-fluid model }

To describe the interaction between two different fluid components, we should introduce two sets of distribution functions.
Each distribution function describes one fluid component, and corresponds to one sets of hydrodynamic quantities (density $\rho^{\sigma}$, flow velocity $\mathbf{u}^{\sigma}$, temperature $T^{\sigma}$, pressure $p^{\sigma}$).
After this step, the two-fluid discrete Boltzmann equation in ES-BGK form can be obtained, i.e.,
\begin{equation}
\frac{\partial f_{i}^{\sigma}}{\partial t}+v_{i\alpha}\cdot\frac{\partial f_{i}^{\sigma}}{\partial r_{\alpha}}=-\frac{1}{\tau^{\sigma}}(f_{i}^{\sigma}-f^{\sigma,ES}_{i})
\label{Eq.Discrete-Boltzmann1}
,
\end{equation}
where the superscript $\sigma$ represents the type of fluid component, i.e., $\sigma$ = A or B.
The $i$ is the kind of discrete velocities.
The relaxation time $\tau^{\sigma}$ relates to the mass particle density $\rho^{\sigma}$ and the flexible parameter $\theta^{\sigma}$, i.e., $\tau^{\sigma}=1/(\rho^{A}/\theta^{\rm{A}}+\rho^{B}/\theta^{\rm{B}})$.
$f^{\sigma,ES}=f^{\sigma,ES}(\rho^{\sigma},\mathbf{u},T)$, with $\rho^{\sigma}$, $\mathbf{u}$, $T$ are the mass density of component $\sigma$, flow velocity of mixture, temperature of mixture, respectively.
In two-fluid DBM, the mass density and flow velocity of each component are calculated by the first two conserved kinetic moments, respectively,  i.e.,
\begin{equation}
\rho^{\sigma}=\sum \nolimits_{i}f^{\sigma}_{i},
\end{equation}
\begin{equation}
\mathbf{u}^{\sigma}=\frac{\sum \nolimits_{i}f^{\sigma}_{i}\mathbf{v}_{i}}{\rho^{\sigma}} ,
\end{equation}
The mass density and flow velocity of mixture are
\begin{equation}
\rho=\sum \nolimits_{\sigma}\rho^{\sigma} ,
\end{equation}
\begin{equation}
\mathbf{u}=\frac{\sum \nolimits_{\sigma}\rho^{\sigma}\mathbf{u}^{\sigma}}{\rho} ,
\end{equation}
The temperatures of each component and the mixture are obtained from the third conserved kinetic moment
\begin{equation}
T^{\sigma*}=\frac{2E_{I}^{\sigma*}}{D \rho^{\sigma}} ,
\end{equation}
\begin{equation}
T=\frac{2E_{I}^{*}}{\sum_{\sigma}D \rho^{\sigma}}
,
\end{equation}
where $E_{I}^{*}=\sum_{\sigma}E_{I}^{\sigma*}$ and $E^{\sigma*}_{I}=\frac{1}{2}\sum_{i}f^{\sigma}_{i}(\mathbf{v}_{i}-\mathbf{u})^2$ is the internal energy of component $\sigma$.
$D$ represents the spatial dimension.

\subsection{ Analysis scheme construction part  }

In the DBM, the non-conserved kinetic moments reflect the manner and extent to which the systems deviate from thermodynamic equilibrium.
By analyzing the non-conservative moments of $(f-f^{eq})$, we can effectively characterize the TNE state and extract valuable TNE information from the fluid system.
Because the DBM used in this paper only considered a limited number of non-conserved kinetic moments, it only captures parts of the TNE behaviors of non-equilibrium systems.
However, these captured TNE behaviors are the most relevant and critical aspects for understanding the system's overall dynamics and characteristics.

In a first-order DBM, four fundamental TNE quantities can be defined, i.e., $\bm{\Delta}^{\sigma*}_{2}$, $\bm{\Delta}^{\sigma*}_{3,1}$, $\bm{\Delta}^{\sigma*}_{3}$, and $\bm{\Delta}^{\sigma*}_{4,2}$.
Their definitions are:
\begin{equation}
\bm{\Delta}^{\sigma*}_{m}=\sum \nolimits_{i}(f_{i}^{\sigma}-f^{\sigma,eq}_{i}) \underbrace { \mathbf{v}^{*}_{i}\mathbf{v}^{*}_{i} \cdots \mathbf{v}^{*}_{i} }_m
, \label{Eq:DDBM-NF28}
\end{equation}
\begin{equation}
\bm{\Delta}^{\sigma*}_{m,n}=\frac{1}{2}\sum \nolimits_{i}(f_{i}^{\sigma}-f^{\sigma,eq}_{i})(\mathbf{v}^{*}_{i}\cdot\mathbf{v}^{*}_{i})^{(m-n)/2} \underbrace { \mathbf{v}^{*}_{i} \cdots \mathbf{v}^{*}_{i} }_n
,
\end{equation}
where $\mathbf{v}^{*}_{i}=\mathbf{v}_{i}-\mathbf{u}$ represents the central velocity, with $\mathbf{u}$ the macro flow velocity.
The subscript ``$m,n$'' means that the $m$-order tensor is contracted to $n$-order tensor.
The first two, $\bm{\Delta}^{\sigma*}_{2}=\Delta^{\sigma*}_{2,\alpha\beta}\mathbf{e}_{\alpha}\mathbf{e}_{\beta}$ and $\bm{\Delta}^{\sigma*}_{3,1}=\Delta^{\sigma*}_{3,1,\alpha}\mathbf{e}_{\alpha}$, are the most typical TNE quantities, where $\mathbf{e}_{\alpha}$ ($\mathbf{e}_{\beta}$) is the unit vector in the $\alpha$ ($\beta$) direction.
Physically, they correspond to more generalized viscous stress (or non-organized momentum flux, NOMF) and heat flux (or non-organized energy flux, NOEF), respectively.
The latter two TNE quantities contain condensed information.
The $\bm{\Delta}^{\sigma*}_{3}$ ($\bm{\Delta}^{\sigma*}_{4,2}$) represents the more generalized flux of $\bm{\Delta}^{\sigma*}_{2}$ ($\bm{\Delta}^{\sigma*}_{3,1}$).
The spatiotemporal evolution of these four TNE quantities is a common scheme that described the way and degree of fluid unit deviating from the equilibrium state.

Mathematically, the expressions for $\Delta^{\sigma*}_{2,\alpha\beta}$ and $\Delta^{\sigma*}_{3,1,\alpha}$ can be found in the corresponding hydrodynamic equations.
Their expressions are
\begin{equation}
\Delta^{\sigma*}_{2,\alpha\beta}=-\mu^{\sigma}(\frac{\partial u^{\sigma}_{\alpha}}{\partial r_{\beta}}+\frac{\partial u^{\sigma}_{\beta}}{\partial r_{\alpha}}-\frac{2}{D}\frac{\partial u^{\sigma}_{\gamma}}{\partial r_{\gamma}}\delta_{\alpha\beta})
,
\label{Eq-delta2*}
\end{equation}
\begin{equation}
\Delta^{\sigma*}_{3,1,\alpha}=-\kappa^{\sigma}\frac{\partial T^{\sigma}}{\partial r_{\alpha}},
\label{Eq-delta31*}
\end{equation}
where $\mu^{\sigma}$ is the dynamic viscosity efficient and $\kappa^{\sigma}$ represents the heat conductivity.
 Please refer to Appendix \ref{sec:AppendixesC} for more derivations.
These two expressions are valuable for grasping the concept of TNE quantity.
However, it's important to note that these two expressions only represent the first order of $\Delta^{\sigma*}_{2,\alpha\beta}$ and $\Delta^{\sigma*}_{3,1,\alpha}$.
TNE quantities defined by DBM constitute a mesoscopic description method originating from non-equilibrium statistical physics.
They hold significant physical interpretations within the phase space.

In addition, other coarse-grained TNE quantities can be defined.
For example, the total TNE strengths describing the extent to which each fluid unit deviates from equilibrium are obtained as follows, i.e.,
\begin{equation}
\left|\bm{\Delta}_{2}^{\sigma *}\right|=\sqrt{\Delta_{2,xx}^{\sigma *2}+2\Delta_{2,xy}^{\sigma *2}+\Delta_{2,yy}^{\sigma *2}}
,
\end{equation}
\begin{equation}
\left|\bm{\Delta}_{3,1}^{\sigma *}\right|=\sqrt{\Delta_{3,1,x}^{\sigma *2}+\Delta_{3,1,y}^{\sigma *2}}
,
\end{equation}
\begin{equation}
\left|\bm{\Delta}_{3}^{\sigma *}\right|=\sqrt{\Delta_{3,xxx}^{\sigma *2}+3\Delta_{3,xxy}^{\sigma *2}
+3\Delta_{3,xyy}^{\sigma *2}+\Delta_{3,yyy}^{\sigma *2}}
,
\end{equation}
\begin{equation}
\left|\bm{\Delta}_{4,2}^{\sigma *}\right|=\sqrt{\Delta_{4,2,xx}^{\sigma *2}+2\Delta_{4,2,xy}^{\sigma *2}
+\Delta_{4,2,yy}^{\sigma *2}}
,
\end{equation}
where the operator ``$| \quad |$'' indicates summing all the components.
Furthermore, by summing their non-dimensional values over the whole fluid field, the global TNE strength described the fluid system can be obtained, i.e., $d_{2}^{\sigma *}=\sum\nolimits_{ix,iy}\left| \bm{\Delta}^{\sigma*2}_{2}/T^2 \right|$, $d_{3}^{\sigma *}=\sum\nolimits_{ix,iy}\left| \bm{\Delta}^{\sigma*2}_{3}/T^3 \right|$, $d_{3,1}^{\sigma *}=\sum\nolimits_{ix,iy}\left| \bm{\Delta}^{\sigma*2}_{3,1}/T^3 \right|$, and $d_{4,2}^{\sigma *}=\sum\nolimits_{ix,iy}\left| \bm{\Delta}^{\sigma*2}_{4,2}/T^4 \right|$, where ``$ix$'' and ``$iy$'' represent the positions of fluid unit.
In addition, by summing the four non-dimensional TNE quantities, another global TNE strength quantity, which contains more condensed information, can also be defined, i.e.,
\begin{equation}
d^{\sigma *}=\sum\nolimits_{ix,iy} \sqrt{ | \frac{\bm{\Delta}^{\sigma*2}_{2}}{T^2} | + | \frac{\bm{\Delta}^{\sigma*2}_{3,1}}{T^3} |  + | \frac{\bm{\Delta}^{\sigma*2}_{3}}{T^3} | + | \frac{\bm{\Delta}^{\sigma*2}_{4,2}}{T^4} | }
.
\end{equation}
For reading, Table \ref{Table2} summarizes the definitions, elements, and physical meanings of the corresponding TNE quantities.
It should be noticed that these TNE quantities describe the TNE features/behaviors from their own perspectives.
The fluid system requires multi-perspectives research.
The results from multi-perspectives together constitute a relatively comprehensive description of the system.
Therefore, in this paper, a non-equilibrium degree/strength/intensity vector $\mathbf{D}$ is introduced, with its elements composed of serval TNE quantities, to provide multiple-perspective non-equilibrium description for the fluid system.
In summary, in this subsection, the DBM provides the most relevant TNE effects which are not convenient to obtain form NS model.

\begin{center}
\begin{table*}[htbp]
\centering
\begin{tabular}{m{2.4cm}<{\centering}m{6cm}<{\centering}m{5cm}<{\centering}m{2.5cm}<{\centering}}
\hline
\hline
TNE quantity & definition & element(s) & physical meaning \\
\hline
$\bm{\Delta}^{\sigma*}_{2}$ & $\sum_{i}(f_{i}^{\sigma}-f^{\sigma,eq}_{i})\mathbf{v}^{*}_{i}\mathbf{v}^{*}_{i}$ & $\Delta^{\sigma*}_{2,xx}$,$\Delta^{\sigma*}_{2,xy}$,$\Delta^{\sigma*}_{2,yy}$ & non-organized momentum flux (NOMF) \\
\hline
$\bm{\Delta}^{\sigma*}_{3,1}$ &
$\frac{1}{2}\sum_{i}(f_{i}^{\sigma}-f^{\sigma,eq}_{i})(\mathbf{v}^{*}_{i}\cdot\mathbf{v}^{*}_{i}+\eta_{i}^{\sigma2})\mathbf{v}^{*}_{i}$ & $\Delta^{\sigma*}_{3,1,x}$,$\Delta^{\sigma*}_{3,1,y}$ & non-organized energy flux (NOEF)  \\
\hline
$\bm{\Delta}^{\sigma*}_{3}$ & $\sum_{i}(f_{i}^{\sigma}-f^{\sigma,eq}_{i})\mathbf{v}^{*}_{i}\mathbf{v}^{*}_{i}\mathbf{v}^{*}_{i}$  & ${\Delta}^{\sigma*}_{3,xxx}$,${\Delta}^{\sigma*}_{3,xxy}$,${\Delta}^{\sigma*}_{3,xyy}$,${\Delta}^{\sigma*}_{3,yyy}$ & the flux of  $\bm{\Delta^{\sigma*}_{2}}$  \\
\hline
$\bm{\Delta}^{\sigma*}_{4,2}$ & $\frac{1}{2}\sum_{i}(f_{i}^{\sigma}-f^{\sigma,eq}_{i})(\mathbf{v}^{*}_{i}\cdot\mathbf{v}^{*}_{i}+\eta_{i}^{\sigma2})\mathbf{v}^{*}_{i}\mathbf{v}^{*}_{i}$ & ${\Delta}^{\sigma*}_{4,2,xx}$,${\Delta}^{\sigma*}_{4,2,xy}$,${\Delta}^{\sigma*}_{4,2,yy}$ & the flux of  $\bm{\Delta^{\sigma*}_{3,1}}$  \\
\hline
$|\bm{\Delta}^{\sigma*}_{2}|$ & $ \sqrt{(\Delta^{\sigma*}_{2,xx})^2+(2\Delta^{\sigma*}_{2,xy})^2+(\Delta^{\sigma*}_{2,yy})^2}$ & scalar & total TNE strength from the view of $\Delta^{\sigma*}_{2}$ \\
\hline
$|\bm{\Delta}^{\sigma*}_{3,1}|$ & $ \sqrt{(\Delta^{\sigma*}_{3,1,x} )^2+( \Delta^{\sigma*}_{3,1,y} )^2}$ & scalar & total TNE strength from the view of $\Delta^{\sigma*}_{3,1}$ \\
\hline
$|\bm{\Delta}^{\sigma*}_{3}|$ & $\sqrt{(\Delta^{\sigma*}_{3,xxx} )^2+(3\Delta^{\sigma*}_{3,xxy} )^2+(3\Delta^{\sigma*}_{3,xyy} )^2+( \Delta^{\sigma*}_{3,yyy} )^2}$ & scalar & total TNE strength from the view of $\Delta^{\sigma*}_{3}$ \\
\hline
$|\bm{\Delta}^{\sigma*}_{4,2}|$ & $\sqrt{(\Delta^{\sigma*}_{4,2,xx})^2+(2\Delta^{\sigma*}_{4,2,xy})^2+(\Delta^{\sigma*}_{4,2,yy})^2}$ & scalar & total TNE strength from view of $\Delta^{\sigma*}_{4,2}$ \\
\hline
$\overline{\Delta_{2,\alpha\beta}^{\sigma *}}$ & $\sum_{ix}\Delta_{2,\alpha\beta}^{\sigma *}/N_x$ & $\overline{\Delta_{2,xx}^{\sigma *}}$,$\overline{\Delta_{2,xy}^{\sigma *}}$,$\overline{\Delta_{2,yy}^{\sigma *}}$ & average TNE strength from the views of $\overline{\Delta_{2,\alpha\beta}^{\sigma *}}$ \\
\hline
$\overline{\Delta_{3,1,\alpha}^{\sigma *}}$ & $\sum_{ix}\Delta_{3,1,\alpha}^{\sigma *}/N_x$ & $\overline{\Delta_{3,1,x}^{\sigma *}}$,$\overline{\Delta_{3,1,y}^{\sigma *}}$ & average TNE strength from the views of $\overline{\Delta_{3,1,\alpha}^{\sigma *}}$ \\
\hline
$d_{2}^{\sigma *}$ & $\sum\nolimits_{ix,iy}\left| \bm{\Delta}^{\sigma*2}_{2}/T^2 \right|$ & scalar & global TNE strength from the view of $d_{2}^{\sigma *}$ \\
\hline
$d_{3}^{\sigma *}$ & $\sum\nolimits_{ix,iy}\left| \bm{\Delta}^{\sigma*2}_{3}/T^3 \right|$ & scalar & global TNE strength from the view of $d_{3}^{\sigma *}$ \\
\hline
$d_{3,1}^{\sigma *}$ & $\sum\nolimits_{ix,iy}\left| \bm{\Delta}^{\sigma*2}_{3,1}/T^3 \right|$ & scalar & global TNE strength from the view of $d_{3,1}^{\sigma *}$ \\
\hline
$d_{4,2}^{\sigma *}$ & $\sum\nolimits_{ix,iy}\left| \bm{\Delta}^{\sigma*2}_{4,2}/T^4 \right|$ & scalar & global TNE strength from the view of $d_{4,2}^{\sigma *}$ \\
\hline
$d^{\sigma *}$ & $\sum\nolimits_{ix,iy} \sqrt{ \left| \frac{\bm{\Delta}^{\sigma*2}_{2}}{T^2} \right| + \left| \frac{\bm{\Delta}^{\sigma*2}_{3,1}}{T^3} \right|  + \left| \frac{\bm{\Delta}^{\sigma*2}_{3}}{T^3} \right| + \left| \frac{\bm{\Delta}^{\sigma*2}_{4,2}}{T^4} \right|} $ & scalar & global TNE strength from the view of $d^{\sigma *}$ \\
\hline
\hline
\end{tabular}
\caption{ The definitions, elements, and physical meanings for the corresponding TNE quantities in DBM, where the operator $\sum_{ix}$ represents summing the TNE quantities for each row of the fluid field and the operator $\sum_{ix,iy}$ indicates summing the TNE quantities for all fluid units.
}
\label{Table2}
\end{table*}
\end{center}

\section{ Numerical simulations and results }\label{Numerical simulations}

\subsection{ Selecting the discrete scheme }

\begin{figure}[htbp]
\center\includegraphics*
[width=0.45\textwidth]{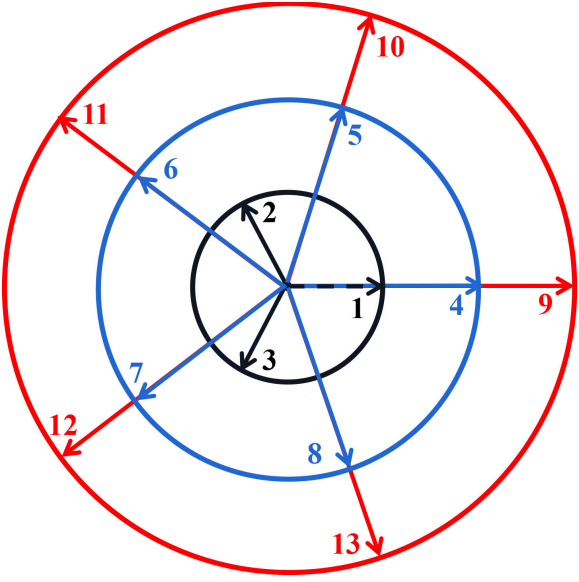}
\caption{Sketch of the D2V13 model.} \label{fig1}
\end{figure}

To obtain the values of $f_{i}^{\sigma,eq}$ and $f_{i}^{\sigma,ES}$, a specific DVM must be selected, i.e., the matrix $\mathbf{C}$ in Eqs. (\ref{Eq.fES}) and (\ref{fieq}).
The selection of DVM must obey the physical constrain imposed by the reserved kinetic moments.
Thus, the dimension of $\mathbf{C}$ is $\mathbf{C}=(\mathbf{c}_1,\mathbf{c}_2,\cdots,\mathbf{c}_{N_{i}})$, with $\mathbf{c}_{i}=(1,v_{ix},v_{iy},\cdots,\frac{1}{2}v_{iy}^2v_{i\alpha}^2)^{T}$, and $N_{i}$ the number of discrete velocities.
The selection of $\mathbf{C}$ also considers the numerical stability and computational efficiency, etc.
Once a specific DVM determined, the values of $f_{i}^{\sigma, eq}$ can be obtained, i.e., $\mathbf{f}^{\sigma, eq}=\mathbf{C}^{-1}\cdot\mathbf{\hat{f}}^{\sigma, eq}$, where $\mathbf{C}^{-1}$ is the inverse matrix of $\mathbf{C}$.
By calculating the values of $\Delta_{2,\alpha \beta}^{\sigma *}$, and submitting they into $\mathbf{\hat{f}}^{\sigma,ES}$, we can get $f_{i}^{\sigma, ES}$, i.e., $\mathbf{f}^{\sigma, ES}=\mathbf{C}^{-1}\cdot\mathbf{\hat{f}}^{\sigma, ES}$.
It should be noticed that solving the inverse matrix is one of the common ways to obtain $\mathbf{f}^{\sigma,eq}$, but it is not the standard or the best method.
In this paper, the number of discrete velocities is chosen to be equal to the number of kinetic moment, i.e., $N_{i}=N_{\rm{moment}}$.
We can use the D2V13 model in this paper, and its sketches can be seen in Fig. \ref{fig1}.
The specific values of D2V13 are given as:
\[
(v_{ix},v_{iy})=
\left\{
\begin{array}{lll}
c_1[\rm{cos} \frac{(\emph{i}-1)2\pi}{3},\rm{sin} \frac{(\emph{i}-1)2\pi}{3}], &i& = 1-3 , \\
c_2[\rm{cos} \frac{(\emph{i}-4)2\pi}{5},\rm{sin} \frac{(\emph{i}-4)2\pi}{5}], &i& = 4-8 , \\
c_3[\rm{cos} \frac{(\emph{i}-9)2\pi}{5},\rm{sin} \frac{(\emph{i}-9)2\pi}{5}], &i& = 9-13 ,
\end{array} \label{Eq:DDBM-DVM}
\right.
\]
where $c_1$, $c_2$ and $c_3$ are adjustable parameters of the DVM.
Then, it is very convenient to obtain $\mathbf{C}^{-1}$ using some mathematical software such as Mathematica, etc.

The standard Lattice Boltzmann Method (LBM) inherits a physical image of ``propagation + collision'' in a given way of ``virtual particle''. \cite{Wangner1998PRL,Swift1995PRL}
In LBM, the direction of discrete velocity represents the motion direction of these ``virtual particle''.
This  concise image is beneficial for improving the computation efficiency of LBM.
However, there is no image of ``propagation + collision'' in DBM.
Although DBM retains the use of discrete velocity, the direction of discrete velocities in DBM does not indicate the motion direction of particles.
The function of DVM in DBM is to keep the values of the reserved kinetic moments.
The specific discrete formats for the spatial derivative, time integral and discrete velocities should be selected reasonably according to the specific situation.
The specific discrete format presented in this paper is only a set of choices based on a series of attempts to meet the current research needs, and is by no means a standard or optimal template.
In this paper, the first-order forward Euler finite difference scheme and the second-order non-oscillatory non-free dissipative scheme are used to solve the two partial derivatives in Eq. (\ref{Eq.Discrete-Boltzmann1}) , respectively.

\subsection{ Configuration and initial conditions }

\begin{figure}[htbp]
\center\includegraphics*
[width=0.5\textwidth]{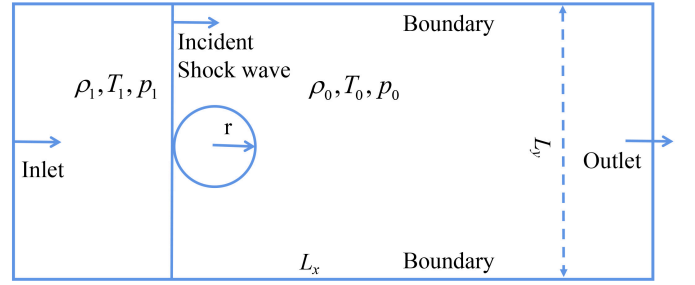}
\caption{The computational configuration of the SBI.}
\label{fig2}
\end{figure}

Figure \ref{fig2} illustrates the configuration of the interaction between a planar shock wave and a cylindrical bubble.
The flow field is rectangular with a dimensionless scale of $L_x \times L_y=0.24 \times 0.12$, where the left side corresponds to the high-pressure region, and the right area is the lower-pressure region.
It is divided into $N_x\times N_y=800 \times 400$ grid size.
The grid number used in the manuscript have passed the grid-independence test.
After the initial moment, a planar shock wave with a strength of $\rm{Ma}=1.23$ propagates downstream and impacts the high-density bubble.
The initial macroscopic conditions of the flow field are as follows:
\[
\left\{ \begin{gathered}
(\rho,T,u_x ,u_y )_{\rm{bubble}} = (5.0168,1.0,0.0,0.0), \hfill \\
(\rho,T,u_x ,u_y )_1 = (1.29201,1.303295,0.393144,0.0), \hfill \\
(\rho,T,u_x ,u_y )_0 = (1.0,1.0,0.0,0.0), \hfill \\
\end{gathered} \right.
\]
where the subscript ``0'' (``1'') represents the low-pressure (high-pressure) region.
Other parameters used for the simulation are: $c_1=0.8$, $c_2=1.6$, $c_3=2.4$, $\Delta x = \Delta y = 3 \times 10^{-4}$, $\Delta t = 1 \times 10^{-6}$, $\theta ^{\rm{A}}=\theta ^{\rm{B}}=4\times10^{-6}$.
The left (right) side of the flow field uses the inflow (outflow) boundary condition.
The periodic boundary is adopted in the $y$ direction.
To study the viscous effects on SBI, we vary the viscosity coefficient of the bubble by adjusting the Pr number.
Fives cases with different $\Pr$ numbers are simulated in this work, i.e., $\Pr=1.0, 1.25, 2.0, 3.33$, and $10.0$.
The larger the $\Pr$ number, the larger the viscosity coefficient (the viscosity formula in ES-BGK DBM can be seen in the Appendix  \ref{sec:AppendixesC}).

\subsection{ Morphological features }

\begin{figure*}[htbp]
\center\includegraphics*
[width=0.6\textwidth]{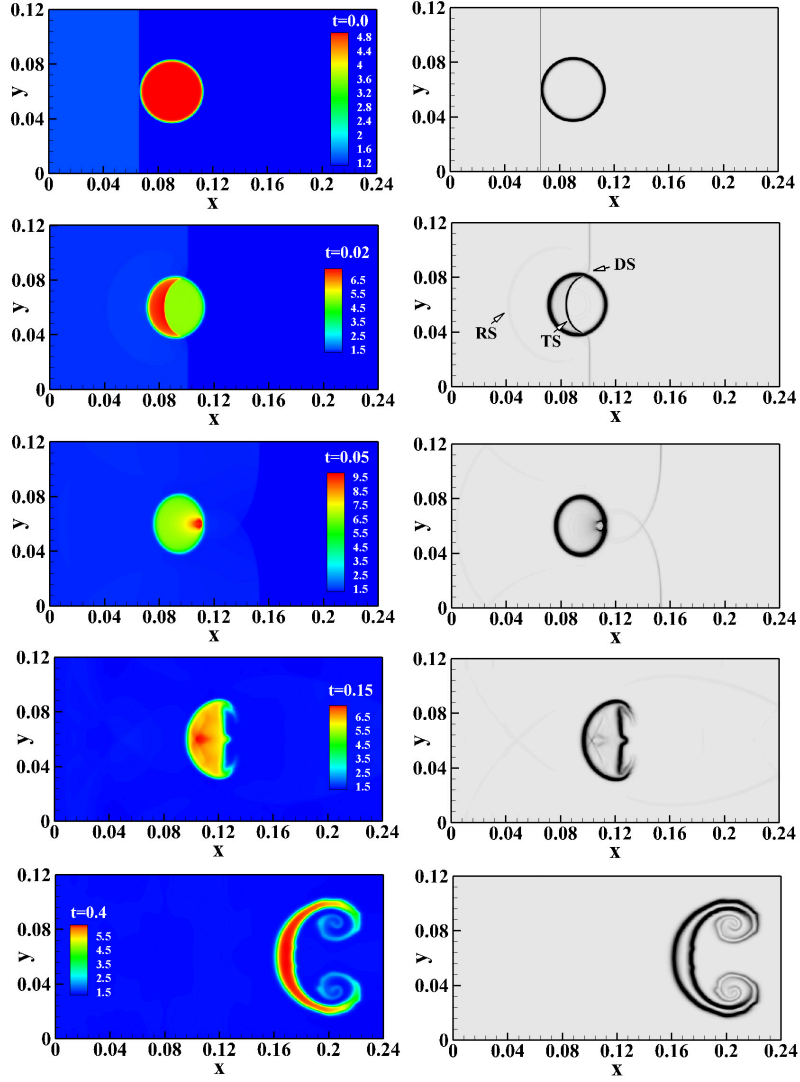}
\caption{Density contours (left column) and schlieren images (right column) at five different moments. }
\label{fig3}
\end{figure*}

\begin{figure*}[htbp]
\center\includegraphics*
[width=0.9\textwidth]{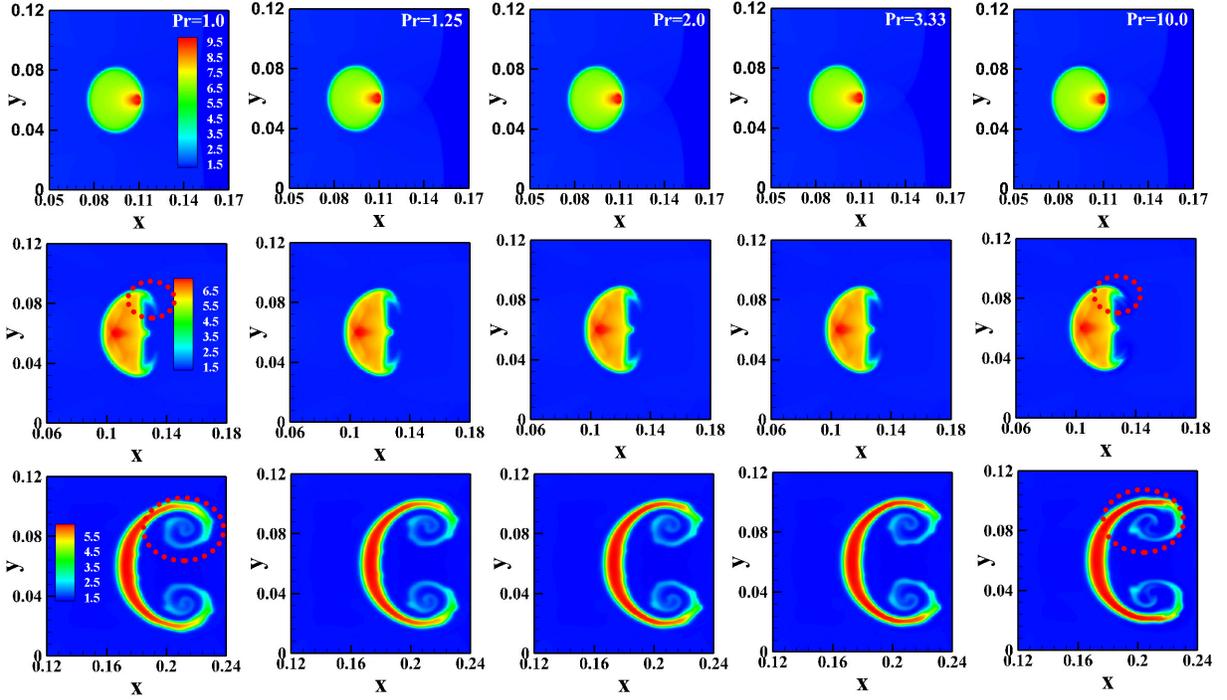}
\caption{Density contours at $t=0.05$ (first row) , $t=0.15$ (second row) and $t=0.42$ (third row), with different $\Pr$ numbers (different viscosities).}
\label{fig4}
\end{figure*}

Shown in Fig. \ref{fig3} are the density contours (left column) and schlieren images (right column) at five different moments, with $\Pr=1.0$.
The SBI can be primarily divided into two stages, i.e., the shock compression stage ($t<0.05$) and the post-shock stage($t>0.05$).
At $t=0$, the incident shock impacts the bubble interface, resulting in the generation of a downstream-traveling transmitted shock (TS) and an upward-moving reflected shock (RS) due to the refraction of the shock wave.
During the shock compression stage, it is characterized by the misalignment between the density gradient and pressure gradient, leading to the vorticity deposition effect which is the intrinsic mechanism behind the formation of two pairs of vortex rings.
Afterward, as TSs move downstream, they converge near the downstream pole, causing the shock focusing and the generation the jet structure.
Subsequently, due to the deposited vorticity, a pair of counter-rotating vortexes is produced.
Figure \ref{fig4} plots the density contours at three moments for cases with five different $\Pr$ numbers (corresponding to different viscosities).
It can be observed that until $t=0.05$, there are almost no differences among cases with different viscosities because the viscous effects are relatively weaker compared to the shock compression effects.
However, during the post-shock stage, the viscous effects are gradually becoming apparent.
The discernible difference (marked by the red circles in the figures) around the undeveloped vortex pair can be observed at $t=0.15$.
 In the case with higher viscosity, the undeveloped vortex pair appears smaller.
 As these vortexes continue to evolve, the influence of viscosity on their shape becomes increasingly pronounced.

\begin{figure*}[htbp]
\center\includegraphics
[width=1.0\textwidth]{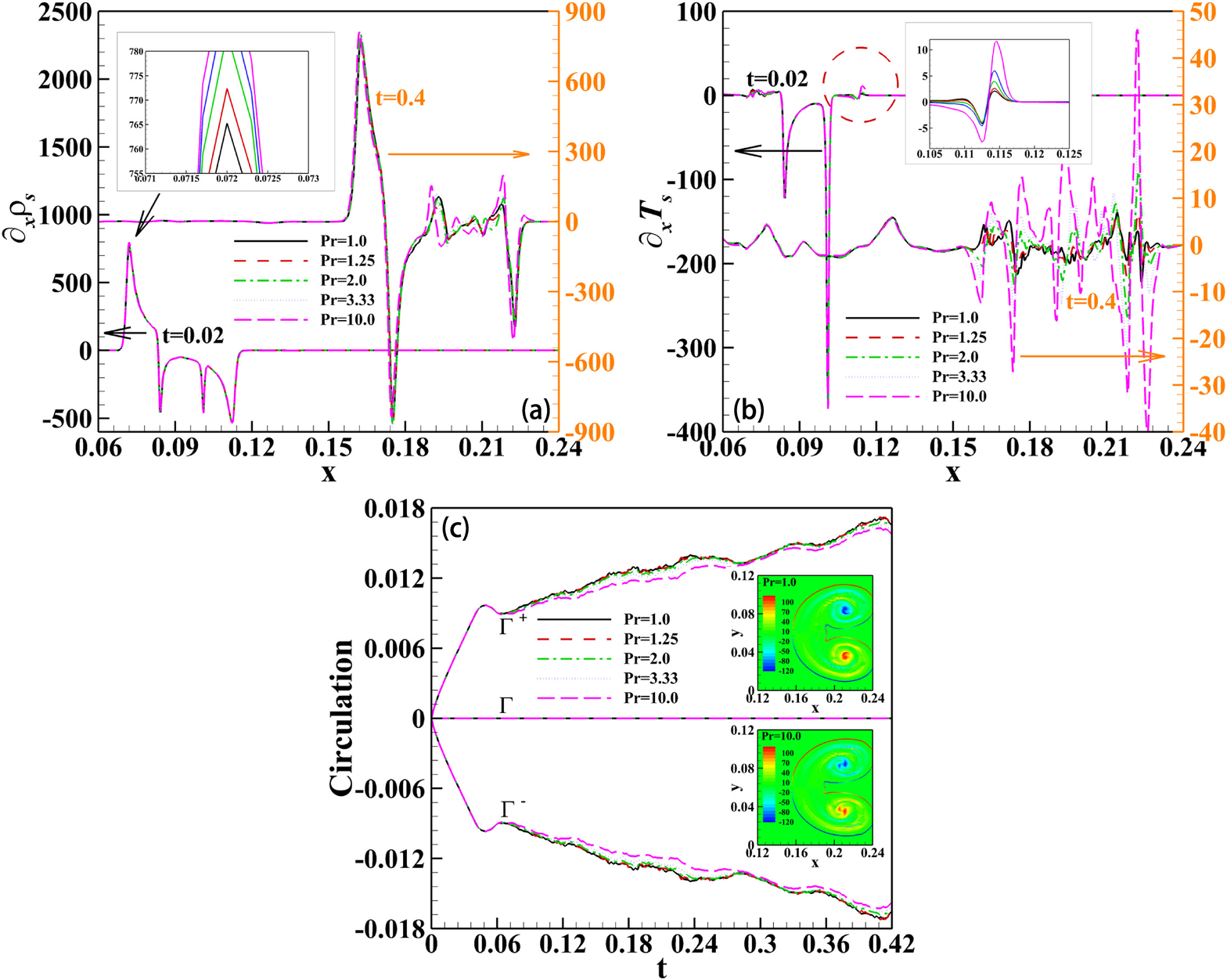}
\caption{(a) Profiles of $\partial_{x} \rho_{s}$ at $t=0.02$ and $t=0.4$, respectively.
(b) Profiles of $\partial_{x} T_{s}$ at $t=0.02$ and $t=0.4$, respectively.
(c) Evolutions of bubble's circulation during the SBI process.
The two sub-figures are the vorticity contours at $t=0.42$.
The upper sub-figure is the case with $\Pr=1.0$ and the bottom one is $\Pr=10.0$.
}
\label{fig41}
\end{figure*}

To further investigate the viscous effects on macroscopic fluid fields, we analyze the density and temperature fields along the $y$ axis by averaging $\rho(ix,iy)$ and $T(ix,iy)$ (i.e., the average density $\rho_s=\sum_{iy}\rho(ix,iy)/N_x$ and average temperature $T_{s}=\sum_{iy}T(ix,iy)/N_x$), and obtain their corresponding gradients (i.e., $\partial_{x} \rho_{s}$ and $\partial_{x} T_{s}$).
As shown in Figs. \ref{fig41}(a) and (b), profiles of $\partial_{x} \rho_{s}$ and $\partial_{x} T_{s}$ at two different moments are plotted, respectively.
At $t=0.02$, there are no obvious differences between cases with different viscosities for both gradients of density and temperature.
However, some subtle but discernible differences can be observed in the corresponding sub-figures.
The case with a higher viscosity exhibits larger peaks.
At $t=0.4$, the differences between cases with different viscosities become more visible.
It also can be seen that the viscosity increases both the amplitudes of density and temperature gradients.
This is because higher-viscosity fluid is more effective at absorbing energy from shock wave.
To investigate the viscous effects on the velocity field, Fig. \ref{fig41}(c) shows the evolution of bubble circulation during the SBI process.
$\Gamma^{+}=\sum \omega|_{\omega >0} \Delta x \Delta y$ ($\Gamma^{-}=\sum \omega|_{\omega < 0} \Delta x \Delta y$) means the positive (negative) circulation and $\Gamma=\sum \omega \Delta x \Delta y$ represents the total circulation, where $\omega = (\partial u_y / \partial x - \partial u_x / \partial y)\mathbf{e}_{z}$ is the vorticity.
The two sub-figures in Fig. \ref{fig41}(c) are the vorticity contours at $t=0.42$.
They show the subtle but discernible differences of vorticity between case $\Pr=1.0$ with case $\Pr=10.0$.
The circulation can be used to describe the strength of vorticity and velocity shear effect.
During the shock compression stage, the circulations increase rapidly due to the vorticity deposition effect.
In this stage, the viscosity contributes little effect to circulations.
However, as the shock wave sweeps through the bubble and the vortex pair continues to develop, the influences of viscosity become  apparent.
Specifically, the viscosity reduces the values of circulation, indicating that the viscosity inhibits the shear motion of the bubble.

The utilization of Minkowski measures is an effective method to extract information from complex physical field. \cite{Gan2011PRE,Gan2019FOP,Xu2009PD,Sofonea1999morphological}
It provides a complete description for a Turing pattern.
In a $D$-dimensional space, a set of convex sets that satisfy motion invariance and additivity can be fully described by $D+1$ Minkowski measures.
In the case of two dimensions, the three Minkowski measures include the proportion $A$ of the high-$\Theta$ region (i.e., $A=A_{h}/A_{total}$, where $A_{h}$ is the area of high-$\Theta$ region and $A_{total}$ is the area of fluid field), the boundary length $L$ between the high- and low-$\Theta$ regions, and the Euler characteristic $\chi$, where $\Theta$ can refer to density, temperature, velocity, or pressure.

\begin{figure*}[htbp]
\center\includegraphics
[width=1.0\textwidth]{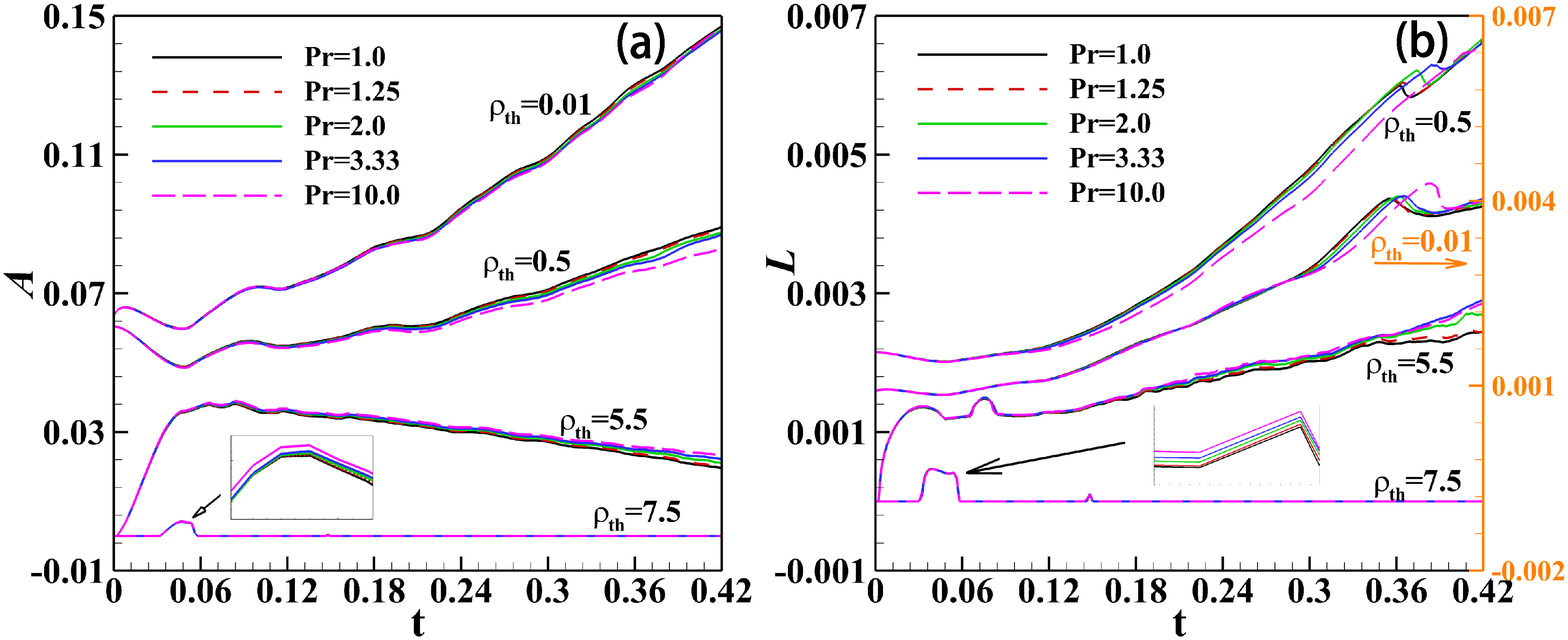}
\caption{(a) Evolutions of proportion $A$.
(b) Evolutions of boundary length $L$.
Four cases with different density threshold values are plotted.
}
\label{fig5}
\end{figure*}

In the following analysis, we focus on density Turing patterns (i.e., $\Theta$ is density).
Fig. \ref{fig5} investigates the viscous effects on the proportion $A$ and boundary length $L$ of density Turing patterns, where ``th'' indicates the ``threshold'' value.
The selection of the threshold value depends on the specific physical effects to be considered.
For example, when $\rho_{th}>\rho_{0}$ (where $\rho_{0}=5.0168$ is the initial density of the bubble), the proportion $A$ and boundary length $L$ reflect the abilities of viscosity to achieve a high-density state due to the shock compression.
When $\rho_{th}<\rho_{0}$, the viscous effects on bubble deformation and diffusion, which result in a lower density, have also been taken into account.
In Fig. \ref{fig5}(a), for cases $\rho_{th}=0.01$ and $\rho_{th}=0.5$, the proportion $A$ decreases rapidly because the bubble is compressed by the incident shock wave.
After the shock wave sweeps through the bubble, the proportion $A$ gradually increases due to the bubble deformation and gas diffusion effects.
During the earlier stage, the viscosity has little impact on the proportion $A$ because the shock compression dominates.
In the later stage, the viscosity slightly reduces the values of proportion $A$ due to its adverse effect on deformation.
When $\rho_{th}=5.5$, the values of proportion $A$ increase from zero due to the shock compression, and then decrease slowly because of the deformation and diffusion.
When further increasing the $\rho_{th}$, the high-density region appears around $t=0.05$ and almost disappears at other times.
In contrast to cases with $\rho_{th}<\rho_{0}$, in cases where $\rho_{th}>\rho_{0}$, the viscosity slightly increases the proportion $A$.
Overall, the viscosity reduces the lower-$\rho_{th}$ region but increases the higher-$\rho_{th}$ region.
The reason is that the fluid with a higher viscosity can absorb more energy from shock wave, making it easier to increase the density during the compression process.
Fig. \ref{fig5} (b) shows the evolutions of boundary length $L$.
Similar to the proportion $A$, the viscous effects are more pronounced in the later stage.
In addition, the viscosity reduces the boundary length $L$ for cases with $\rho_{th}=0.01$ and $\rho_{th}=0.5$, but increases it for  cases with $\rho_{th}=5.5$ and $\rho_{th}=7.5$.

\subsection{ TNE features }

\begin{figure*}[htbp]
\center\includegraphics
[width=0.8\textwidth]{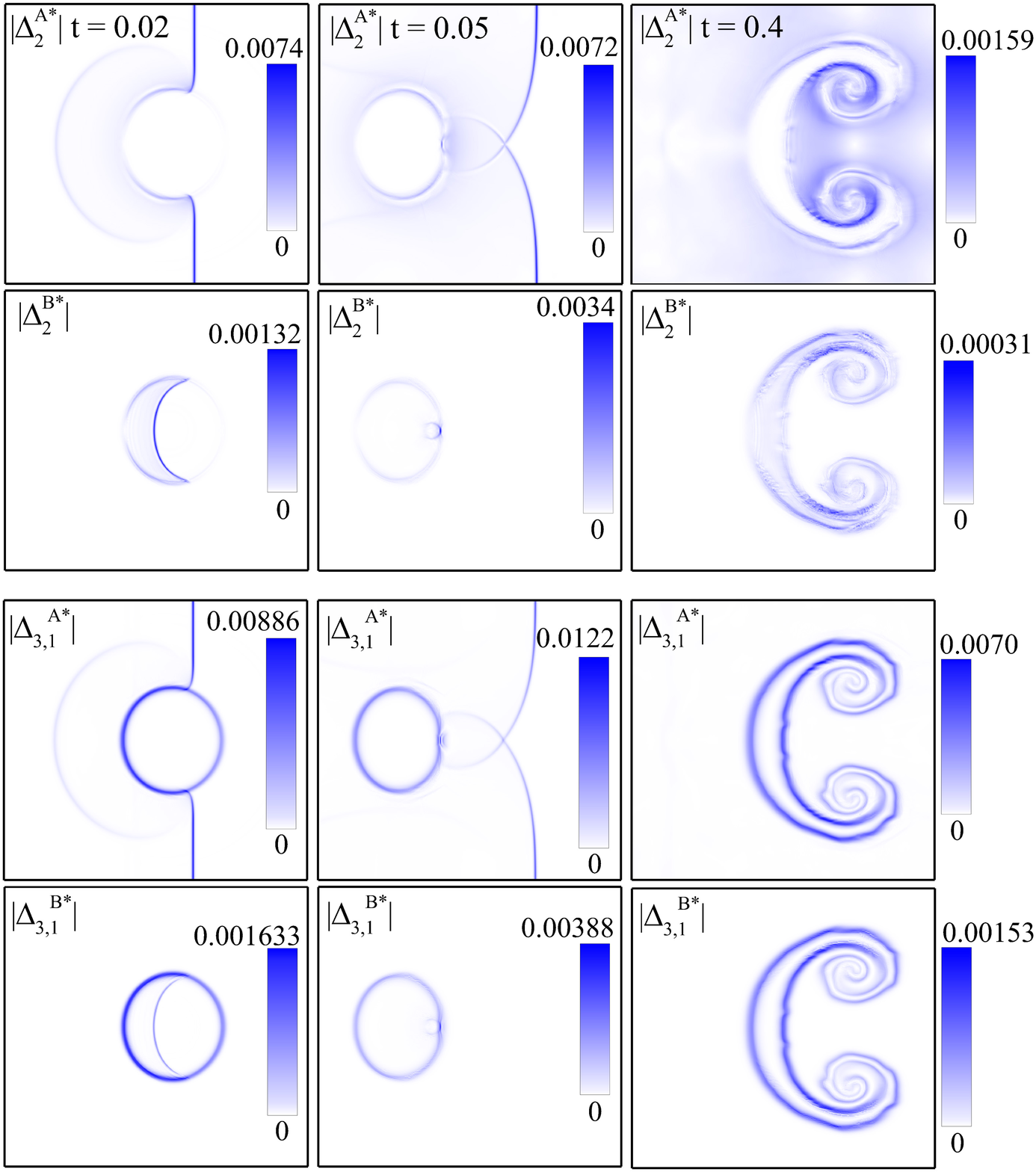}
\caption{ Contours of $|\bm{\Delta}_{2}^{\sigma,*}|$ and $|\bm{\Delta}_{3,1}^{\sigma,*}|$ at three different moments.
}
\label{fig6}
\end{figure*}

Understanding the spatial distribution and temporal evolution of TNE quantities is crucial for further investigating the kinetic behaviors during the SBI process.
In the following part, we introduce a non-equilibrium strength vector $\mathbf{D}$, in which its elements are composed of $\{ |\bm{\Delta}_{2}^{\sigma*}|, |\bm{\Delta}_{3,1}^{\sigma*}|, \overline{\Delta_{2,\alpha\beta}^{\sigma *}}, \overline{\Delta_{3,1,\alpha}^{\sigma *}}, d^{\sigma*}_{2}, d^{\sigma*}_{3,1}, d^{\sigma*} \}$, to analyze the TNE strength of fluid system from multiple perspectives.
To provide an intuitive representation of TNE quantities, Fig. \ref{fig6} shows the contours of the total TNE strengths at three different moments.
In this figure, the odd and even rows represent the quantities of component $A$ and $B$, and the first and last two rows correspond the total TNE strengths $|\bm{\Delta}_{2}^{\sigma*}|$ and $|\bm{\Delta}_{3,1}^{\sigma*}|$, respectively.
It can be observed that the values of total TNE strengths are greater than zero in regions where the gradients of macroscopic quantities are pronounced, indicating significant deviations from the thermodynamic equilibrium state.
The larger the value of the TNE quantity, the higher the degree of deviation from the thermodynamic equilibrium state.

\begin{figure*}[htbp]
\center\includegraphics
[width=0.9\textwidth]{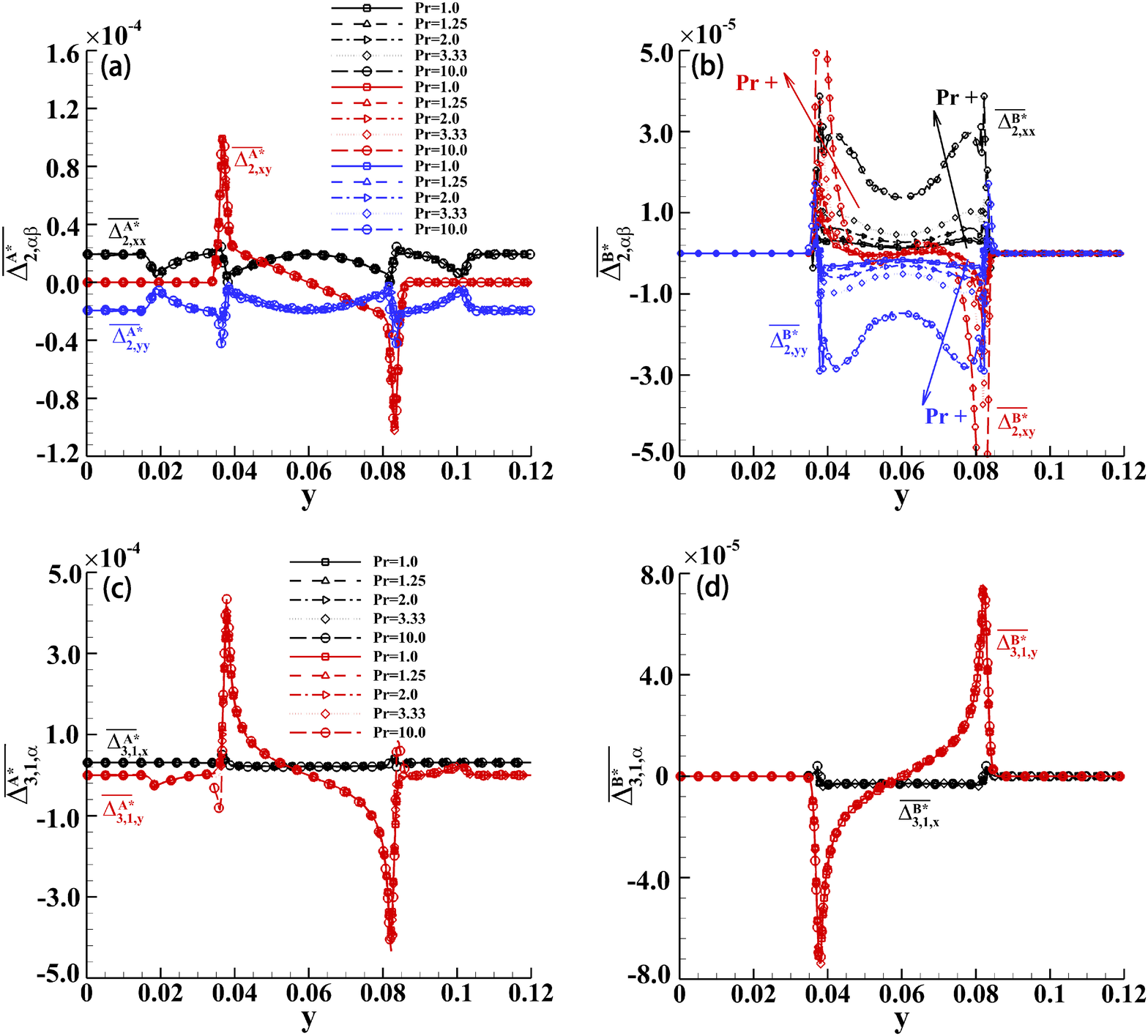}
\caption{ (a) Profiles of $\overline{\Delta_{2,\alpha\beta}^{A*}}$.
(b) Profiles of $\overline{\Delta_{2,\alpha\beta}^{B*}}$.
(c) Profiles of $\overline{\Delta_{3,1,\alpha}^{A*}}$.
(d) Profiles of $\overline{\Delta_{3,1,\alpha}^{B*}}$.
}
\label{fig7}
\end{figure*}

To qualitatively investigate the viscous effects on the spatial distribution of TNE quantities, two average TNE strengths quantities are analyzed.
Shown in Figs. \ref{fig7}(a)-(d) are the profiles of average TNE strengths $\overline{\Delta_{2,\alpha\beta}^{A *}}$, $\overline{\Delta_{2,\alpha\beta}^{B *}}$, $\overline{\Delta_{3,1,\alpha}^{A *}}$, and $\overline{\Delta_{3,1,\alpha}^{B *}}$ at moment $t=0.02$, respectively.
Different symbols in the figure indicate cases with various viscosities.
It is evident that the values of average TNE strengths around the contact interface are relatively larger, and the profiles of these TNE quantities show obvious symmetry.
Actually, these TNE quantities describe the TNE strength from their own perspectives.
Specifically, Fig. \ref{fig7}(a) shows the views from $\overline{\Delta_{2,\alpha\beta}^{A *}}$.
Clearly, the profiles of $\overline{\Delta_{2,xx}^{A *}}$ are symmetric about the central line $y=0.6$, indicating that the upper and lower parts of the fluid field deviate from the thermodynamic equilibrium state in the same direction and with the same amplitude.
When viewing the TNE strength from the perspective of $\overline{\Delta_{2,yy}^{A *}}$, the results are the same as above because the profiles of $\overline{\Delta_{2,yy}^{A *}}$ are also symmetric about the central line $y=0.6$.
In addition, profiles of $\overline{\Delta_{2,xx}^{A *}}$ and $\overline{\Delta_{2,yy}^{A *}}$ are symmetric about the lines $\overline{\Delta_{2,\alpha\beta}^{A*}}=0$, which means that looking at the TNE strength from these two perspectives yields opposite results.
What can also be seen is that the profiles of the average TNE strength $\overline{\Delta_{2,xy}^{A *}}$ are symmetry about the origin point.
It indicates when looking at the view from $\overline{\Delta_{2,xy}^{A *}}$, the upper and lower parts of the flow field deviate from equilibrium in opposite directions but deviating with the same amplitude.
As shown in Fig. \ref{fig7}(b), profiles of $\overline{\Delta_{2,xx}^{B *}}$ and $\overline{\Delta_{2,yy}^{B *}}$ are also symmetric about the lines $\overline{\Delta_{2,\alpha\beta}^{B*}}=0$.
The profiles of $\overline{\Delta_{2,xy}^{B *}}$ are also symmetry about the origin point.

In addition, it can be seen that the viscosity does not have too many influences on $\overline{\Delta_{2,\alpha\beta}^{A *}}$, except some differences around the contact interface.
However, the viscosity increases obviously the values of $\overline{\Delta_{2,\alpha\beta}^{B *}}$, indicating that the viscosity enhances the deviation from the thermodynamic equilibrium state.
The reason for this can be understood by examining how the $\Pr$ number affects the TNE quantity.
For understanding, we can refer to the Eq. (\ref{Eq-delta2*}).
Theoretically, viscosity affects the TNE quantities through influencing the transport coefficient and gradients of macroscopic quantity.
When raising the viscosity of component $B$, the transport coefficients of $\overline{\Delta_{2,\alpha\beta}^{B *}}$ increase, resulting in the growth of strengths of $\overline{\Delta_{2,\alpha\beta}^{B *}}$.
However, the viscosity changes the strengths of $\overline{\Delta_{2,\alpha\beta}^{A *}}$ by affecting the gradients of macroscopic quantity around the contact interface between two fluids.
When changing the viscosity of component $B$, in the early time of SBI, the influences on gradients of macroscopic quantities of component $A$ are weak.
As time goes, the viscous effects on macroscopic quantity gradients gradually become more apparent.

Figures \ref{fig7}(c) and (d) show the average TNE strength from the perspectives of $\overline{\Delta_{3,1,\alpha}^{\sigma *}}$.
Compared to the strengths of $\overline{\Delta_{3,1,x}^{\sigma *}}$, the strengths of $\overline{\Delta_{3,1,y}^{\sigma *}}$ exhibit more significant amplitude, indicating that the average heat flux in $y$ direction is stronger than that in $x$ direction.
The profiles of $\overline{\Delta_{3,1,x}^{\sigma *}}$ are symmetric about the line $y=0.6$ and profiles of $\overline{\Delta_{3,1,y}^{\sigma *}}$ are symmetric about the origin point.
The $\Pr$ number has limited effects on strengths of $\overline{\Delta_{3,1,x}^{\sigma *}}$ and $\overline{\Delta_{3,1,y}^{\sigma *}}$ since the viscosity does not directly change their transport coefficients ( as shown in Eq. (\ref{Eq-delta31*}) ).

There are two points that should be stated: (i) It is essential to note that the above descriptions are based on the results at a typical moment $t=0.02$.
Descriptions of other moments are equally important for a comprehensive understanding of the kinetic behaviors during the SBI process.
(ii) Moreover, higher-order TNE quantities ($\Delta_{3,\alpha\beta\gamma}^{\sigma*}$ and $\Delta_{4,2,\alpha\beta}^{\sigma*}$) contain more condensed information.
They are also closely related to the gradients of macroscopic quantities.

\begin{figure*}[htbp]
\center\includegraphics
[width=0.9\textwidth]{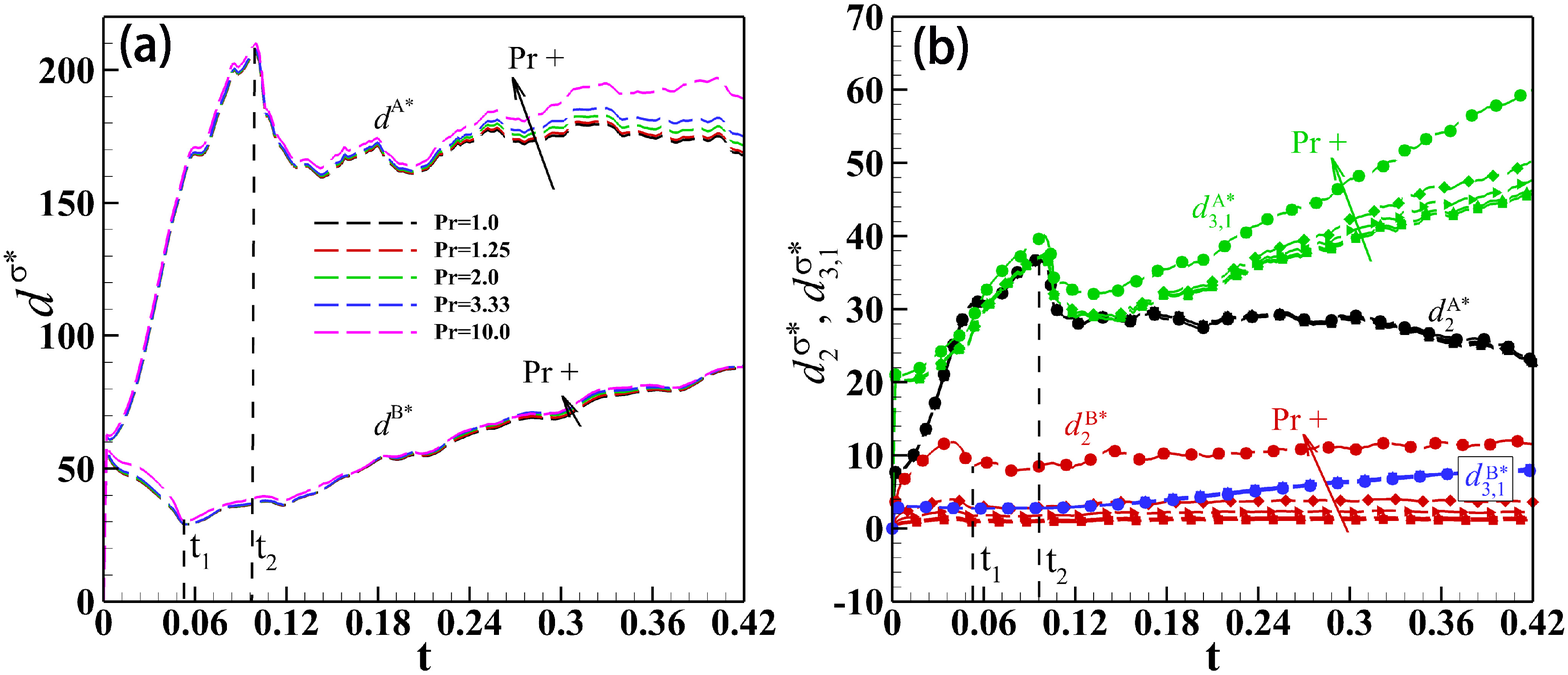}
\caption{
(a) Evolutions of global TNE strengths $d^{A *}$ and $d^{B *}$.
(b) Evolutions of global TNE strengths $d_{2}^{\sigma *}$ and $d_{3,1}^{\sigma *}$.
}
\label{fig8}
\end{figure*}

To qualitatively investigate the viscous effects on TNE strength of fluid system, Fig. \ref{fig8}(a) plots the evolutions of the global TNE strength $d^{\sigma *}$.
It can be seen that the peaks (valleys) of $d^{A *}$ and $d^{B*}$ profiles are closely related to the location of the shock wave.
Here the $t_{1}$ indicates the moment that the incident shock has just passed the bubble, and the $t_{2}$ represents the moment when the incident shock exits the fluid field.
It is also shown that the viscosity increases both the global TNE strengths of the two components.
Further, Fig. \ref{fig8}(b) shows the evolutions of global TNE strengths $d_{2}^{\sigma *}$ and $d_{3,1}^{\sigma *}$.
Lines with different colors indicate cases with different TNE components, and different symbols represent different viscosities.
For component $B$ (the bubble), the viscosity significantly enhances the strength of $d_{2}^{B*}$ but not changes the strength of $d_{3,1}^{B*}$.
The reason for this is that when adjusting the $\Pr$ number, it does not directly modify the transport coefficient of $d_{3,1}^{B*}$.
For component $A$ (the ambient gas), the viscosity significantly increases the strength of $d_{3,1}^{A*}$ but not affects the strength of $d_{2}^{A*}$.
The reason is attributed to the fact that a fluid with higher viscosity exhibits enhanced capacity for absorbing energy from shock waves, thereby resulting in an elevated heat flux.

\begin{figure*}[htbp]
\center\includegraphics
[width=0.9\textwidth]{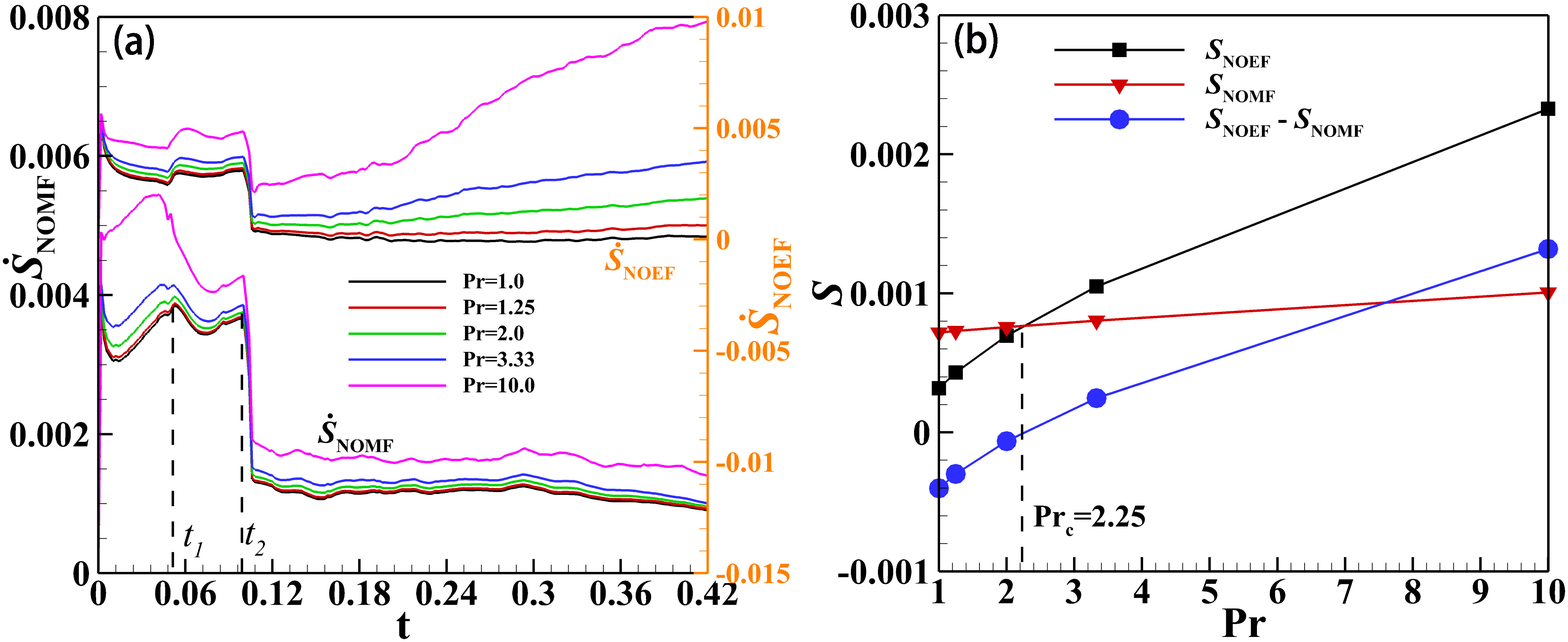}
\caption{ (a) Evolutions of entropy production rates $\dot{S}_{\rm{NOEF}} $ and $\dot{S}_{\rm{NOMF}}$.
(b) Entropy productions ($S_{\rm{NOEF}}$, $S_{\rm{NOMF}}$, and $S_{\rm{NOEF}}$ - $S_{\rm{NOMF}}$) over this period.
}
\label{fig10}
\end{figure*}

The entropy production rate and entropy production, which are significant in the compression science field, are also analyzed.
There are two kinds of entropy production rates \cite{Zhang2019Matter}:
\begin{equation}
\dot{S}_{\rm{NOEF}} =  \int \bm{\Delta}_{3,1}^{*} \cdot \nabla \frac{1}{T} d \bm{r}
,
\end{equation}
\begin{equation}
\dot{S}_{\rm{NOMF}} =  \int -\frac{1}{T} \bm{\Delta}_{2}^{*} : \nabla \bm{u} d \bm{r}
.
\end{equation}
The former is caused by the NOEF and the temperature gradient, and the latter is contributed by the NOMF and the velocity gradient.
Fig. \ref{fig10}(a) plots the evolutions of entropy production rates $\dot{S}_{\rm{NOEF}}$ and $\dot{S}_{\rm{NOMF}}$.
It can be seen that the changes in the curves of the entropy generation rate are closely related to the position of the shock wave.
For $\dot{S}_{\rm{NOMF}}$, its values increase continuously during the process when the incident shock wave is acting on the bubble ($t<t_{1}$).
Later, the values of $\dot{S}_{\rm{NOMF}}$ decrease.
When the incident shock wave runs out of the fluid field ($t>t_{2}$), the $\dot{S}_{\rm{NOMF}}$ reduces rapidly and subsequently maintain their values for a longer time.
For $\dot{S}_{\rm{NOEF}}$, before $t_{2}$, its values reduce and then increase.
When $t>t_{2}$, the $\dot{S}_{\rm{NOEF}}$ shows an upward trend.
The viscosity significantly raises the $\dot{S}_{\rm{NOMF}}$ by increasing the transport coefficient, and amplifies the  $\dot{S}_{\rm{NOEF}}$ by increasing the temperature gradient.
Summing the entropy generation rates over this period, the entropy generations ($S_{\rm{NOEF}}$ and $S_{\rm{NOMF}}$) during this period can be obtained.
Fig. \ref{fig10}(b) shows the profiles of entropy generations versus viscosities ($\Pr$ numbers).
It can be seen that both the two types of entropy generations increase as the viscosity increases.
Before $\Pr<\Pr_{c}$, the $S_{\rm{NOMF}}$ is larger than $S_{\rm{NOEF}}$.
When $\Pr>\Pr_{c}$, the situation is reversed.

\section{Conclusions}\label{Conclusions}

A two-fluid DBM with a flexible $\Pr$ number is designed to investigate the influence of viscosity on the morphological and TNE characterizations during the SBI process.
Different from most of the previous numerical research that relied on traditional macroscopic models/methods, this paper studies the dynamic and kinetic processes of SBI from a mesoscopic view.
Overall, for the rapid SBI process, the viscosity contributes subtle but discernible influences on macroscopic quantities while strongly affects the some TNE features of the fluid system.
Morphologically, (i) the viscosity affects the shape of the vortex pair, increases both the amplitudes of gradients of density and temperature of the fluid field, reduces the values of the circulation of the bubble.
(ii) the viscosity increases both the proportion of the area occupied by the high-density region and the boundary length separating the high- and low-density regions.
The underlying reason is that the fluid with higher viscosity possesses an enhanced ability to absorb energy from shock waves, thus facilitating an increase in density during the compression process.

Introducing a non-equilibrium strength vector $\mathbf{D}$, defined as $\mathbf{D} = \{ |\bm{\Delta}_{2}^{\sigma*}|, |\bm{\Delta}_{3,1}^{\sigma*}|, \overline{\Delta_{2,\alpha\beta}^{\sigma *}}, \overline{\Delta_{3,1,\alpha}^{\sigma *}}, d^{\sigma*}_{2}, d^{\sigma*}_{3,1}, d^{\sigma*} \}$, offers diverse non-equilibrium descriptions from multiple perspectives for the fluid system.
The spatial distributions of TNE quantities demonstrate that the contact interfaces with macroscopic quantity gradients significantly deviate from thermodynamic equilibrium, while the regions far from the interface remain close to the equilibrium state.
The spatial configuration of average TNE strength show interesting symmetry.
Different TNE quantities describe the TNE strength from their corresponding perspectives, and the results from various perspectives together constitute a relatively comprehensive description of the system.
Theoretically, the viscosity influences these TNE quantities by affecting the transport coefficient and gradients of macroscopic quantities.
Through raising the transport coefficient and the temperature gradient, respectively, the viscosity increases the two types of entropy production rates and their corresponding entropy production.
The fundamental research presented in this paper enhances our understanding of the SBI mechanism in various applications, such as inertial confinement fusion, supersonic combustors, underwater explosions, etc.

\section*{Acknowledgments}
The authors thank Chuandong Lin, Feng Chen, Ge Zhang, Yiming Shan, Jie Chen and Hanwei Li on helpful discussions for DBM.
This work was supported by the National Natural Science Foundation of China ( Nos.  12172061, 11875001, and 12102397), the Strategic Priority Research Program of Chinese Academy of Sciences ( No. XDA25051000), the opening project of State Key Laboratory of Explosion Science and Technology (Beijing Institute of Technology) ( No. KFJJ23-02M), Foundation of National Key Laboratory of Computational Physics, Foundation of National Key Laboratory of Shock Wave and Detonation Physics (No. JCKYS2023212003), and Hebei Natural Science Foundation ( Nos. A2021409001 and A2023409003), Central Guidance on Local Science and Technology Development Fund of Hebei Province ( No. 226Z7601G), and ``Three, Three and Three Talent Project'' of Hebei Province ( No. A202105005).

\appendix

\section{ Expressions of the kinetic moments of $f^{\sigma,ES}$ } \label{sec:AppendixesB}
The kinetic moments can be obtained by integrating the Eq. (\ref{Eq.fES}) in the particle velocity space $\bf{v}$.
Their expressions are as follows:
\begin{equation}
M^{\sigma,ES}_{0}=\sum_{i}f^{\sigma,ES}_{i}=\rho^{\sigma} ,
\label{M0}
\end{equation}
\begin{equation}
M^{\sigma,ES}_{1,x}=\sum_{i}f^{\sigma,ES}_{i}v_{ix}=\rho^{\sigma}u_{x} ,
\end{equation}
\begin{equation}
M^{\sigma,ES}_{1,y}=\sum_{i}f^{\sigma,ES}_{i}v_{iy}=\rho^{\sigma}u_{y} ,
\end{equation}
\begin{equation}
M^{\sigma,ES}_{2,xy}=\sum_{i}f^{\sigma,ES}_{i}v_{ix}v_{iy}=\rho^{\sigma}(\lambda_{xy}+u_{x}u_{y}) ,
\end{equation}
\begin{equation}
M^{\sigma,ES}_{2,xx}=\sum_{i}f^{\sigma,ES}_{i}v_{ix}^{2}=\rho^{\sigma}(\lambda_{xx}+u_{x}^2)  ,
\end{equation}
\begin{equation}
M^{\sigma,ES}_{2,yy}=\sum_{i}f^{\sigma,ES}_{i}v_{iy}^{2}=\rho^{\sigma}(\lambda_{yy}+u_{y}^2)  ,
\end{equation}
\begin{equation}
M^{\sigma,ES}_{3,xxx}=\sum_{i}f^{\sigma,ES}_{i}v_{ix}^{3}=\rho^{\sigma}u_{x}(3\lambda_{xx}+u_{x}^{2}) ,
\end{equation}
\begin{equation}
M^{\sigma,ES}_{3,xxy}=\sum_{i}f^{\sigma,ES}_{i}v_{ix}^{2}v_{iy}=\rho^{\sigma}[2\lambda_{xy}u_x+u_y(\lambda_{xx}+u_x^2)] ,
\end{equation}
\begin{equation}
M^{\sigma,ES}_{3,xyy}=\sum_{i}f^{\sigma,ES}_{i}v_{ix}v_{iy}^{2}=\rho^{\sigma}u_{x}[2\lambda_{xy}u_y+u_x(\lambda_{yy}+u_y^2)] ,
\end{equation}
\begin{equation}
M^{\sigma,ES}_{3,yyy}=\sum_{i}f^{\sigma,ES}_{i}v_{iy}^{3}=\rho^{\sigma}u_{y}(3\lambda_{yy}+u_{y}^{2}) ,
\end{equation}
\begin{eqnarray}
\begin{aligned}
M^{\sigma,ES}_{4,2,xx}&=\sum_{i}f^{\sigma,ES}_{i}\frac{1}{2}v_{ix}^2v_{i\alpha}^{2}=\frac{1}{2} \rho^{\sigma} [\lambda_{xx} (\lambda_{yy}+6 u_{x}^{2}+u_{y}^{2})\\
&+3 \lambda_{xx}^{2}+4 \lambda_{xy} u_x u_y+2 \lambda_{xy}^{2}+u_x^2 (\lambda_{yy}+u_x^2+u_y^2)]
,
\end{aligned}
\end{eqnarray}
\begin{eqnarray}
\begin{aligned}
M^{\sigma,ES}_{4,2,xy}&=\sum_{i}f^{\sigma,ES}_{i}\frac{1}{2}v_{ix}v_{iy}v_{i\alpha}^{2}=\frac{1}{2}\rho^{\sigma} [3 \lambda_{xx} (\lambda_{xy}+u_x u_y)\\
&+3 \lambda_{xy} (\lambda_{yy}+u_x^2+u_y^2)+u_x u_y (3 \lambda_{yy}+u_x^2+u_y^2)]
,
\end{aligned}
\end{eqnarray}
\begin{eqnarray}
\begin{aligned}
M^{\sigma,ES}_{4,2,yy}&=\sum_{i}f^{\sigma,ES}_{i}\frac{1}{2}v_{iy}^2v_{i\alpha}^{2}=\frac{1}{2}\rho^{\sigma} [u_y^2 (\lambda_{xx}+6 \lambda_{yy}+u_x^2)\\
&+\lambda_{yy} (\lambda_{xx}+3 \lambda_{yy}+u_x^2)+4 \lambda_{xy} u_x u_y+2 \lambda_{xy}^2+u_y^4]
,
\label{M42yy}
\end{aligned}
\end{eqnarray}
The modified term $\lambda_{\alpha\beta}=R^{\sigma}T\delta_{\alpha\beta}+\frac{b^{\sigma}}{\rho^{\sigma}}\Delta^{\sigma *}_{2,\alpha\beta}$ with $\Delta_{2,\alpha \beta}^{\sigma *}=\sum_{i}(f_{i}^{\sigma}-f^{\sigma,eq}_{i})\mathbf{v}^{*}_{i}\mathbf{v}^{*}_{i}$.
The two contracted moments can be expressed by the corresponding non-contracted moments, i.e.,
\begin{equation}
M^{\sigma,ES}_{2,0}=\frac{1}{2}(M^{\sigma,ES}_{2,xx}+M^{\sigma,ES}_{2,yy})
\end{equation}
\begin{equation}
M^{\sigma,ES}_{3,1,x}=\frac{1}{2}(M^{\sigma,ES}_{3,xxx}+M^{\sigma,ES}_{3,xyy})
\end{equation}
\begin{equation}
M^{\sigma,ES}_{3,1,y}=\frac{1}{2}(M^{\sigma,ES}_{3,yyy}+M^{\sigma,ES}_{3,xxy})
\end{equation}
When $b^{\sigma}=0$, the ES-BGK model is simplified to BGK model.
So the kinetic moments of $f^{\sigma,eq}$ can be obtained easily by submitting $b^{\sigma}=0$ into the kinetic moments of $f^{\sigma,ES}$.

\section{Two-fluid hydrodynamic equations} \label{sec:AppendixesC}

In this part, the expressions of hydrodynamic equations derived from the ES-BGK Boltamann equation through the CE multiscale analysis are given.
More detailed derivations can see the reference presented by Zhang \emph{et al.} \cite{Zhang2020POF}.
The NS equations for component $\sigma$ are:
\begin{equation}
\frac{\partial\rho^{\sigma}}{\partial t}+\frac{\partial}{\partial r_{\alpha}}(\rho^{\sigma} u_{\alpha}^{\sigma})=0
,
\end{equation}
\begin{eqnarray}
\begin{aligned}
\frac{\partial}{\partial t}(\rho^{\sigma} u_{\alpha}^{\sigma})&+\frac{\partial (p^{\sigma}\delta_{\alpha\beta}+\rho^{\sigma}
u^{\sigma}_{\alpha}u^{\sigma}_{\beta})}{\partial r_{\beta}}+\frac{\partial (P^{\sigma}_{\alpha\beta}+U_{\alpha \beta}^{\sigma})}{\partial r_{\beta}}\\
&=-\frac{\rho^{\sigma}}{\tau^{\sigma}}(u_{\alpha}^{\sigma}-u_{\alpha}), \label{Eq:DDBM-N1}
\end{aligned}
\end{eqnarray}
\begin{eqnarray}
\begin{aligned}
&\frac{\partial}{\partial t}(\rho^{\sigma} E_{T}^{\sigma})+\frac{\partial}{\partial r_{\alpha}}(\rho^{\sigma}E^{\sigma}_{T}+p^{\sigma})u^{\sigma}_{\alpha}+\frac{\partial}{\partial r_{\beta}}[u^{\sigma}_{\beta}(P^{\sigma}_{\alpha\beta}+U_{\alpha\beta}^{\sigma})\\
&-\kappa^{\sigma}\frac{\partial T^{\sigma}}{\partial r_{\alpha}}+Y_{\alpha}^{\sigma}]=-\frac{\rho^{\sigma}}{\tau^{\sigma}}[\frac{D(T^{\sigma}-T)}{2}+\frac{1}{2}(u_{\alpha}^{\sigma 2}-u_{\alpha}^{2})] .
\end{aligned}
\end{eqnarray}
where
\begin{equation}
E_{T}^{\sigma}=\frac{1}{2}(DT^{\sigma}+u_{\alpha}^{\sigma 2})
,
\end{equation}
\begin{equation}
P^{\sigma}_{\alpha\beta}=-\mu^{\sigma}(\frac{\partial u^{\sigma}_{\alpha}}{\partial r_{\beta}}+\frac{\partial u^{\sigma}_{\beta}}{\partial r_{\alpha}}-\frac{2}{D}\frac{\partial u^{\sigma}_{\gamma}}{\partial r_{\gamma}}\delta_{\alpha\beta})
,
\end{equation}
\begin{equation}
U_{\alpha\beta}^{\sigma}=\rho^{\sigma}[(u_{\beta}-u^{\sigma}_{\beta})(u_{\alpha}-u^{\sigma}_{\alpha})
+\frac{1}{D}(u^{\sigma}_{\alpha}-u_{\alpha})^{2}\delta_{\alpha\beta}]
,
\end{equation}
\begin{eqnarray}
\begin{aligned}
Y_{\alpha}^{\sigma}&=[\frac{D+2}{2}\rho^{\sigma}R^{\sigma}(T^{\sigma}-T)(u^{\sigma}_{\alpha}-u_{\alpha})
-\frac{D+4}{2D}\rho^{\sigma}(u_{\alpha}^{\sigma}-u_{\alpha})^2u_{\alpha}^{\sigma}\\
&+\rho^{\sigma}u_{\alpha}^{\sigma}
(u_{\alpha}^{\sigma}-u_{\alpha})u_{\alpha}^{\sigma}-\rho^{\sigma}u_{\beta}^{\sigma}(u_{\alpha}u_{\beta}-u_{\alpha}u_{\beta}^{\sigma})]\\
&+\frac{1}{2}\rho^{\sigma}(u^{2}_{\alpha}-u_{\alpha}^{\sigma 2})u_{\alpha}
.
\end{aligned}
\end{eqnarray}

Performing $\sum_{\sigma}$ on both sides of the three equations gives the NS hydrodynamic equations describing the whole system, i.e.,
\begin{equation}
\frac{\partial\rho}{\partial t}+\frac{\partial}{\partial r_{\alpha}}(\rho u_{\alpha})=0
,
\end{equation}
\begin{eqnarray}
\begin{aligned}
\frac{\partial}{\partial t}(\rho u_{\alpha})
&+\frac{\partial \sum\limits_{\sigma}(p^{\sigma}\delta_{\alpha\beta}+\rho^{\sigma}
u^{\sigma}_{\alpha}u^{\sigma}_{\beta})}{\partial r_{\beta}}\\
&+\frac{\partial\sum\limits_{\sigma} (P^{\sigma}_{\alpha\beta}+U_{\alpha \beta}^{\sigma})}{\partial r_{\beta}}=0, \label{Eq:DDBM-N1}
\end{aligned}
\end{eqnarray}
\begin{eqnarray}
\begin{aligned}
&\frac{\partial}{\partial t}(\rho E_{T})+\frac{\partial}{\partial r_{\alpha}}\sum\limits_{\sigma}(\rho^{\sigma}E^{\sigma}_{T}+p^{\sigma})u^{\sigma}_{\alpha}\\
&+\frac{\partial}{\partial r_{\beta}}\sum\limits_{\sigma}[u^{\sigma}_{\beta}(P^{\sigma}_{\alpha\beta}+U_{\alpha\beta}^{\sigma})-\kappa^{\sigma}\frac{\partial T^{\sigma}}{\partial r_{\alpha}}+Y_{\alpha}^{\sigma}]=0 .
\end{aligned}
\end{eqnarray}

The dynamic viscosity efficient is $\mu^{\sigma}=\Pr^{\sigma} \tau^{\sigma} p^{\sigma}=\frac{1}{1-b^{\sigma}}\tau^{\sigma} p^{\sigma}$, with $p^{\sigma}=\rho^{\sigma} R^{\sigma} T^{\sigma}$.
The heat conductivity is $\kappa^{\sigma}=C_{p}^{\sigma}\tau^{\sigma} p^{\sigma}$, with $C_{p}^{\sigma}=\frac{D+2}{2}R^{\sigma}$.
It should be pointed out again that recovering the hydrodynamic equations is only one part of the physical function of DBM.
The physical function of DBM corresponds to the EHEs.
In DBM, it does not need to solve the hydrodynamic equations in DBM simulation.

\section*{Data Availability}
The data that support the findings of this study are available from the corresponding author upon reasonable request.

\section*{References}
\bibliography{SBI-POF}

\end{document}